\documentclass[journal]{IEEEtran}
\usepackage{times} 		% assumes new font selection scheme installed
\usepackage{mathtools,amssymb,mathrsfs} 	% for math, includes amsmath package
\usepackage{xcolor} 	% for coloured texts, includes color package
\usepackage{cite} 		% hyphen replacing consecutive numbers
\usepackage{verbatim} 	% for commenting out blocks
\usepackage{graphicx} 	% for figures, includes graphics package
\usepackage{epsfig} 	% for postscript graphics files
\usepackage[per-mode=symbol]{siunitx}	% for SI units
\usepackage{stfloats} 		% for bottom long equation
\usepackage{float} 		% for figure right [H]ere
\newtheorem{ass}{Assumption}
\newtheorem{theorem}{Theorem}
\newtheorem{rem}{Remark}

\newtheorem{pro}{Proposition}
\newtheorem{cor}{Corollary}
\newtheorem{exmp}{Example}

\DeclareMathOperator{\tri}{tri}
\allowdisplaybreaks[1]

\definecolor{note_fontcolor}{rgb}{0.800781, 0.800781, 0.800781}

\makeatother

\pdfoutput=1

\begin{document}
	
	\title{Adaptive Scan for Atomic Force Microscopy Based on Online Optimisation:
		Theory and Experiment}
	
	\markboth{IEEE Transactions on Control Systems Technology}{Wang \MakeLowercase{\textit{et al.}}: Adaptive Scan for Atomic
		Force Microscopy Based on Online Optimisation: Theory and Experiment}
	
	\author{Kaixiang~Wang, Michael~G.~Ruppert, Chris~Manzie, Dragan~Ne{\v{s}}i{\'{c}},
		and Yuen~K.~Yong\thanks{Kaixiang Wang is with the Department of Mechanical Engineering, The
			University of Melbourne, Parkville, VIC 3010, Australia. (e-mail:
			kaixiangw@student.unimelb.edu.au).}\thanks{Michael~G.~Ruppert and Yuen~K.~Yong are with the School of Electrical
			Engineering and Computing, The University of Newcastle, Callaghan,
			NSW 2308, Australia (e-mail: michael.ruppert@newcastle.edu.au; yuenkuan.yong@newcastle.edu.au).}\thanks{Chris Manzie and Dragan Ne{\v{s}}i{\'{c}} are with the Department
			of Electrical and Electronic Engineering, The University of Melbourne,
			Parkville, VIC 3010, Australia. (e-mail: manziec@unimelb.edu.au; dnesic@unimelb.edu.au).} }
	\maketitle
	\begin{abstract}
		A major challenge in Atomic Force Microscopy (AFM) is to reduce the
		scan duration while retaining the image quality. Conventionally, the
		scan rate is restricted to a sufficiently small value in order to
		ensure a desirable image quality as well as a safe tip-sample contact
		force. This usually results in a conservative scan rate for samples
		that have a large variation in aspect ratio and/or for scan patterns
		that have a varying linear velocity. In this paper, an adaptive scan
		scheme is proposed to alleviate this problem. A scan line-based performance
		metric balancing both imaging speed and accuracy is proposed, and
		the scan rate is adapted such that the metric is optimised online
		in the presence of aspect ratio and/or linear velocity variations.
		The online optimisation is achieved using an extremum-seeking (ES)
		approach, and a semi-global practical asymptotic stability (SGPAS)
		result is shown for the overall system. Finally, the proposed scheme
		is demonstrated via both simulation and experiment. 
	\end{abstract}
	
	
	\begin{comment}
	Standard Phrases:
	
	scan rate\\
	imaging speed, scan speed\\
	imaging duration, scan duration\\
	imaging accuracy, image quality\\
	set point piezoelectric actuator, piezo actuator 
	\end{comment}
	
	\section{Introduction}
	
	\label{sec:INTRO}
	
	Atomic Force Microscopy (AFM) is used to image nano-scale samples
	such as biological and chemical structures \cite{Santos2004,Gross2009}.
	The instrument works by scanning a cantilever over the sample surface
	laterally while maintaining the vertical interaction force between
	the cantilever tip and the sample. In order to obtain an accurate
	estimate of the sample topography, a vertical actuation and feedback
	system is typically in place to regulate the tip-sample interaction
	force to some set point, and the displacement of the actuator is then
	inverted to obtain the sample topography image \cite{Abramovitch2007}.
	
	There is a desire to increase the AFM imaging speed in order to achieve
	a higher throughput, ultimately leading to video-rate imaging. However,
	this is limited by the XY scanning system and vertical force regulation
	system\cite{Fleming2010c}. The XY scanning system typically drives
	piezoelectric actuators to track the XY reference trajectories which
	form a scan pattern that suitably covers the scan area. Due to the
	inherent nonlinearities in piezoelectric actuators such as creep and
	hysteresis, feedback methods are typically implemented to ensure a
	good tracking performance \cite{Sebastian2005,Leang2007,Esbrook2013a}.
	However, the closed-loop bandwidth is limited by the lightly damped
	resonances of the actuators, which in turn limits the bandwidth of
	the scan \cite{Das2014}. This is exacerbated by the use of conventional
	raster scan, since the triangular trajectory possesses a high waveform
	bandwidth \cite{Devasia2007}.
	
	To alleviate this problem, various non-raster scan patterns that are
	composed of low-bandwidth scan trajectories have been proposed \cite{Yong2010c,Mahmood2011,Bazaei2012a}.
	Relevant output regulation controllers (sometimes referred to as internal
	model controllers or servocompensators) are designed to further increase
	the tracking performance \cite{Yong2014,Bazaei2016}. Additionally,
	combined feedforward/feedback schemes have been proposed to overcome
	the limitations in feedback controller design \cite{Butterworth2012,Kara-Mohamed2015}.
	
	In addition to the XY scanning system, the vertical regulation system
	also limits the imaging speed. In order to obtain an accurate topography
	estimate, the regulation error needs to be small. However, the scanning
	motion and the varying sample topography in effect generate a disturbance
	to the system whose signal bandwidth increases with the scan rate.
	Therefore, for a given regulation system, the only way to improve
	the imaging accuracy is to decrease the scan rate \cite{Wang2016}.
	
	One way to overcome this limitation is to increase the Z feedback
	loop bandwidth through novel actuator designs and/or high-bandwidth
	robust controller designs \cite{Fleming2011,Kuiper2012a,Yong2013c,Schitter2001}.
	Alternatively, the underlying connection between XY and Z systems
	can be utilised. The scan is usually conducted in a line-by-line manner
	with two consecutive scan lines very close to each other. This implies
	that the sample topography signal along two consecutive lines would
	be similar. Several researchers have attempted to utilise this characteristic
	and devised iterative, repetitive or feedforward controllers to improve
	the vertical regulation response \cite{Ren2014,Necipoglu2011a,Schitter2004c}.
	
	All aforementioned methods assume a fixed scan rate. This may lead
	to a conservative imaging speed for samples that have a large variation
	in aspect ratio and/or scan patterns that have a varying linear velocity.
	The scan rate needs to be sufficiently small to ensure a small regulation
	error and safe contact force even in the worst case \cite{Wang2016}.
	Therefore, it is desirable to adapt the scan rate to these variations
	in order to achieve a faster imaging speed while retaining a desirable
	image quality.
	
	A few rule-based adaptive scan methods have been proposed in \cite{Heuell1994,Ahmad2014,Ren2016}.
	However, these methods react to the instantaneous regulation error
	and are not concerned with the optimality of scan. In \cite{Wang2017},
	we have proposed a novel adaptive scan scheme based on the online
	optimisation of a performance metric. Representing the trade-off between
	imaging speed and accuracy, the metric can be a user-specified function
	of the scan rate and regulation error over each line to suit particular
	applications. As one potential choice, the performance metric can
	penalise the deviation of the maximum regulation error over a line
	from some set point. In this way, it is expected that a better image
	can be obtained under a certain duration, thereby improving the imaging
	speed-accuracy trade-off. Since the sample topography is unknown,
	the optimal scan rate that optimises the metric is in general unknown.
	Therefore, in \cite{Wang2017}, we have implemented the extremum-seeking
	(ES) framework in \cite{Haring2013} to search for the unknown optimal
	scan rate. Due to sample topography variations, the optimal scan rate
	is time-varying. However, since two consecutive scan lines are close
	to each other, the regulation error along the two lines are similar.
	Therefore, it is expected that the optimal scan rate is slowly varying
	across scan lines. In \cite{Wang2017}, we have shown via simulation
	that the algorithm may track the optimal scan rate approximately and
	retain a semi-global practical asymptotic stability (SGPAS).
	
	This paper extends our preliminary work presented in \cite{Wang2017}
	both theoretically and experimentally. In this paper, a rigorous theoretical
	proof is given for the SGPAS result, which in itself is a nontrivial
	extension of that in \cite{Haring2013} due to the consideration of
	a time-varying cost function. Secondly, a systematic comparison between
	the proposed adaptation law and the existing adaptation law in \cite{Ren2016}
	is performed, and indicates superior performance of the proposed method.
	Thirdly, this paper discusses a practical scenario where the adaptive
	scan scheme can be used to improve the imaging performance. Finally,
	the adaptive scan scheme is implemented on a commercial AFM platform,
	demonstrating the efficacy of the proposed approach.
	
	The rest of this paper is organized as follows. In Section \ref{sec:PF},
	the AFM system is introduced and the performance function is defined.
	In Section \ref{sec:ES}, the adaptation law is described based on
	the design of an ES controller that optimises the performance function,
	and the SGPAS result is shown. The simulation and experiment results
	are presented in Section \ref{sec:SR} and \ref{sec:ER}, respectively.
	Finally, conclusions are drawn in Section \ref{sec:CON}.
	
	\section{Problem Formulation}
	
	\label{sec:PF}
	
	In this section, the AFM system is introduced and the performance
	function is defined. A few mathematical assumptions are posed in order
	to facilitate forthcoming theoretical performance analysis.
	
	\subsection{System Model}
	
	\begin{figure}[!t]
		\centering \includegraphics[width=8.4cm]{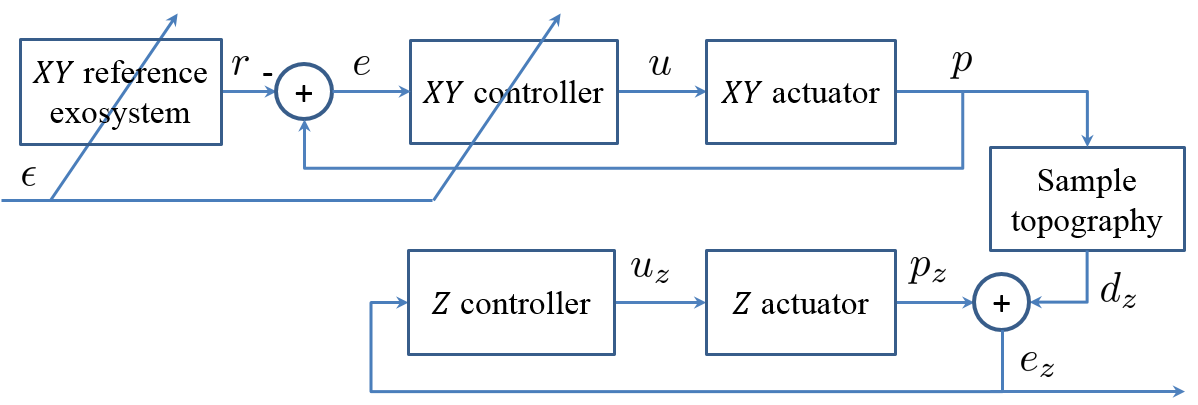}
		\caption{AFM system block diagram.}
		\label{fig:overall_block_diagram} 
	\end{figure}
	
	\begin{figure}[!t]
		\centering \includegraphics[width=8.4cm]{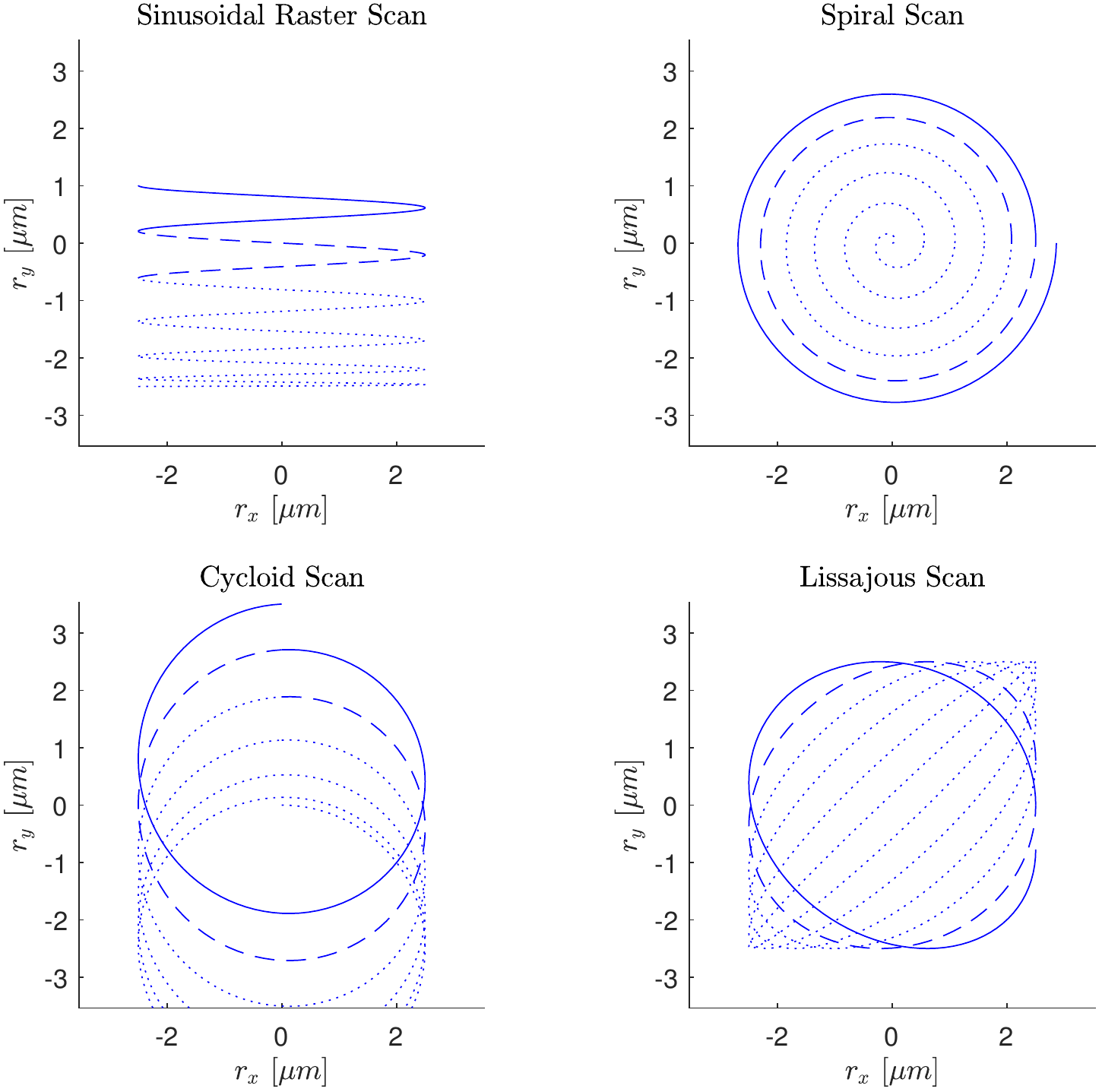}
		\caption{Definition of scan lines. The solid and dashed lines represent a set
			of current and previous scan lines, respectively. }
		\label{fig:scan_line_definition} 
	\end{figure}
	
	Consider a typical AFM system structure illustrated in Fig. \ref{fig:overall_block_diagram}.
	$r\in\mathbb{R}^{2}$ denotes the XY reference trajectories, $u\in\mathbb{R}^{2}$
	the XY control inputs, $p\in\mathbb{R}^{2}$ the XY actuator displacements
	and $e\in\mathbb{R}^{2}$ the XY tracking error. The sample topography
	relates the XY actuator displacements to the sample topography signal
	$d_{z}\in\mathbb{R}$, which acts as an output disturbance to the
	vertical feedback system. $u_{z}\in\mathbb{R}$, $p_{z}\in\mathbb{R}$
	and $e_{z}\in\mathbb{R}$ represent the control input, actuator displacement,
	and regulation error along the Z axis, respectively. The XY and Z
	controllers are designed to minimise $e$ and $e_{z}$, respectively.
	
	\begin{comment}
	The XY controller is designed to drive the XY actuator to track the
	XY reference trajectories, $r$, with minimal the XY tracking error,
	$e$, while the Z controller is designed to drive the Z actuator to
	reject the disturbance, $d_{z}$, to reduce the Z regulation error,
	$e_{z}$. 
	\end{comment}
	
	A range of scan patterns are available in the literature: conventional
	raster or sinusoidal raster scan, cycloid scan \cite{Yong2010c},
	spiral scan \cite{Mahmood2011}, and Lissajous scan \cite{Bazaei2012a}.
	All these scan methods are conducted in a line-by-line manner and
	each line is referred to as a \emph{scan line} (see Fig. \ref{fig:scan_line_definition}).
	The nominal duration for completing a scan line is referred to as
	the \emph{nominal line period} and is denoted by $T_{l}$. In practice,
	it is usually possible to modify the \emph{number of scan lines},
	$N$, which would affect the \emph{nominal scan duration}, $T_{0}$.
	Moreover, in order to trade off between speed and accuracy, it is
	also possible to vary the duration for completing a scan line, or
	the \emph{line period}, by modifying the \emph{scan rate multiplier},
	$\epsilon$. Both $N$ and $\epsilon$ affect the scan duration.
	
	\begin{ass} \label{ass:reference} The XY references trajectories
		can be generated by a linear neutrally stable exosystem model:
		
		\begin{equation}
		\begin{aligned}\dot{\xi}_{r} & =\epsilon A_{r}\left(N\right)\xi_{r}\\
		r & =C_{r}\xi_{r}
		\end{aligned}
		,\label{eq:xyr}
		\end{equation}
		where $\xi_{r}\in D_{r}\subset\mathbb{R}^{n_{r}}$ denotes the exosystem
		state, and $A_{r}\left(N\right)$ and $C_{r}$ are matrices that determine
		the shape of the scan path. Notice that the dependence of $A_{r}$
		on $N$ reflects the fact that the shape of the scan path depends
		on the number of scan lines. 
	\end{ass} 

	\begin{rem}
		All aforementioned scan patterns have reference trajectories that
		can be cast into or approximated by the above form (\ref{eq:xyr})
		with appropriate initial conditions \cite{Wang2016}. It should be
		noted that although the triangular trajectory may be present in some
		of the scan patterns, they can be approximated by a finite sum of
		sinusoids whose frequency is the odd harmonics of the fundamental
		frequency \cite{Lathi2009}, which is then compatible with (\ref{eq:xyr}).
		This approximation can simplify subsequent analysis, but is not required
		for the implementation of the proposed scheme in practice, as will
		be shown in the simulation result section. 
	\end{rem}

	Since $\epsilon$ could be time-varying, to facilitate the forthcoming
	analysis, an auxiliary timescale $\tau$ is defined by 
	\begin{equation}
	\dot{\tau}=\epsilon\left(t\right)\label{eq:tau}
	\end{equation}
	with $\tau\left(0\right)=0$. It can be seen that this definition
	reduces to $\tau=\epsilon t$ for a constant $\epsilon$, and that
	$\tau\left(t\right)$ is invertible as long as $\epsilon\left(t\right)$
	is positive for all $t\geq0$.
	
	\begin{rem}
		\label{rem:tau-timescale} $\tau$ can be seen as the distance, rather
		than the duration, that the scan has progressed. This should be clear
		by rewriting (\ref{eq:xyr}) in the $\tau$-timescale,
		\begin{equation}
		\frac{d\xi_{r}}{d\tau}=A_{r}\left(N\right)\xi_{r},\label{eq:xyr2}
		\end{equation}
		and noticing its solution does not depend on $\epsilon$. %
		\begin{comment}
		This also implies that the line period in the $\tau$-timescale is
		$T_{l}$ uniformly in $\epsilon$. 
		\end{comment}
	\end{rem}

	Since the forthcoming analysis will mainly be done in the $\tau$-timescale,
	the following abuse of notation is used to simplify the presentation.
	For an arbitrary signal $y\left(t\right)$ and its trajectory in the
	$\tau$-timescale defined by $y_{\tau}\left(\tau\right)\triangleq y\left(t\left(\tau\right)\right)$,
	the subscript will be dropped when there is no confusion, i.e. $y_{\tau}\left(\tau\right)$
	is simply written as $y\left(\tau\right)$. 
	
	Now, two examples of the reference trajectories are provided below
	given the scan size\footnote{The side length of a square scan area.}
	$L$, the nominal line period $T_{l}$ and the number of scan lines
	$N$. Also define $A_{l}=\frac{L}{2}$, $\omega_{l}=\frac{2\pi}{T_{l}}$, $\omega_{fr}=\frac{2\pi}{\left(2N-1\right)T_{l}}$ and $\omega_{fs}=\frac{2\pi}{4NT_{l}}$.
	
	\begin{exmp} \label{exmp:sra} The sinusoidal raster scan reference
		trajectories can be generated by 
		\begin{align*}
		r_{x}\left(\tau\right) & =A_{l}\sin\left(\omega_{l}\tau-\frac{\pi}{2}\right),\\
		r_{y}\left(\tau\right) & =A_{l}\sin\left(\omega_{fr}\tau-\frac{\pi}{2}\right),
		\end{align*}
		which can be realised by a model in the form of (\ref{eq:xyr2}) with
		\begin{align*}
		A_{r} & =\begin{bmatrix}0 & \omega_{l} & 0 & 0\\
		-\omega_{l} & 0 & 0 & 0\\
		0 & 0 & 0 & \omega_{fr}\\
		0 & 0 & -\omega_{fr} & 0
		\end{bmatrix},\ \xi_{r}\left(0\right)=\begin{bmatrix}-A_{l}\\
		0\\
		-A_{l}\\
		0
		\end{bmatrix},\\
		C_{r} & =\begin{bmatrix}1 & 0 & 0 & 0\\
		0 & 0 & 1 & 0
		\end{bmatrix}.
		\end{align*}
	\end{exmp}
	
	\begin{exmp} \label{exmp:spi} Likewise, the spiral scan reference
		trajectories are\footnote{The proposed spiral scan trajectories are different from those in
			\cite{Mahmood2011,Bazaei2016} in the sense that they are sinusoidally
			modulated rather than triangularly modulated. } 
		\begin{align*}
		r_{x}\left(\tau\right) & =A_{l}\sin\left(\omega_{fs}\tau\right)\cos\left(\omega_{l}\tau\right),\\
		r_{y}\left(\tau\right) & =A_{l}\sin\left(\omega_{fs}\tau\right)\sin\left(\omega_{l}\tau\right).
		\end{align*}
		Using the trigonometric
		identity, they are equivalent to 
		\begin{align*}
		r_{x}\left(\tau\right) & =\frac{1}{2}A_{l}\left(\sin\left(\left(\omega_{l}+\omega_{fs}\right)\tau\right)-\sin\left(\left(\omega_{l}-\omega_{fs}\right)\tau\right)\right),\\
		r_{y}\left(\tau\right) & =\frac{1}{2}A_{l}\left(\cos\left(\left(\omega_{l}-\omega_{fs}\right)\tau\right)-\cos\left(\left(\omega_{l}+\omega_{fs}\right)\tau\right)\right),
		\end{align*}
		which can be similarly generated using (\ref{eq:xyr2}). \end{exmp}
	
	The following assumption is placed on the sample topography to facilitate
	forthcoming analysis. 
	
	\begin{ass} \label{ass:topography} The sample topography is represented
		by a static map, $f_{s}:\mathbb{R}^{2}\rightarrow\mathbb{R}$, which
		is unknown but is sufficiently many times differentiable. \end{ass} 
	\begin{rem}
		In practice, the sample can have vertical edges or even overhanging
		features. However, the shape and size of the cantilever tip as well
		as the tip cone angle inevitably produce some apparent slope. Therefore,
		despite the apparent limiting nature of this mathematical assumption,
		from a practical viewpoint it is not a significant impediment. 
	\end{rem}
	Consider the XY actuator 
	\begin{equation}
	\begin{aligned}\dot{\xi}_{o} & =A_{o}\xi_{o}+B_{o}u\\
	p & =C_{o}\xi_{o}\\
	e & =p-r
	\end{aligned}
	,\label{eq:xyo}
	\end{equation}
	the XY controller 
	\begin{equation}
	\begin{aligned}\dot{\xi}_{c} & =A_{c}\left(\epsilon,N\right)\xi_{c}+B_{c}\left(\epsilon,N\right)e\\
	u & =C_{c}\left(\epsilon,N\right)\xi_{c}
	\end{aligned}
	,\label{eq:xyc}
	\end{equation}
	the Z actuator 
	\begin{equation}
	\begin{aligned}\dot{\xi}_{zo} & =A_{zo}\xi_{zo}+B_{zo}u_{z}\\
	p_{z} & =C_{zo}\xi_{zo}\\
	e_{z} & =p_{z}+f_{s}\left(p\right)
	\end{aligned}
	,\label{eq:zo}
	\end{equation}
	and the Z controller 
	\begin{equation}
	\begin{aligned}\dot{\xi}_{zc} & =A_{zc}\xi_{zc}+B_{zc}e_{z}\\
	u_{z} & =C_{zc}\xi_{zc}
	\end{aligned}
	,\label{eq:zc}
	\end{equation}
	where $\xi_{o}\in\mathbb{R}^{n_{o}}$ and $\xi_{c}\in\mathbb{R}^{n_{c}}$
	represent the XY actuator and controller state variables, respectively,
	and $\xi_{zo}\in\mathbb{R}^{n_{zo}}$, $\xi_{zc}\in\mathbb{R}^{n_{zc}}$
	the Z counterparts. 
	\begin{rem}
		It is common to model the actuators with linear models preceded by
		hysteresis nonlinearity \cite{Croft2001,Yong2012,Gu2013}. Although
		the form of (\ref{eq:xyo}) and (\ref{eq:zo}) here neglects the hysteresis
		effect, it is assumed that the hysteresis can be cancelled by means
		of hysteresis inversion \cite{Croft2001} or charge drives \cite{Fleming2005}. 
	\end{rem}
	\begin{rem}
		The XY controller (\ref{eq:xyc}) and Z controller (\ref{eq:zc})
		are stated in a general form to cover a wide range of existing control
		algorithms on these axes with the adaptive scan augmentation. Their
		designs are not the intended contribution of this paper. The dependence
		of (\ref{eq:xyc}) on $\epsilon$ and $N$ is in order to permit asymptotic
		tracking. A similar construct appears in \cite{Esbrook2013}, where
		the argument of the system matrices is the online estimate of a presumably
		unknown frequency for the exogenous reference and disturbance. 
	\end{rem}
	\begin{exmp} \label{exmp:3}In \cite{Esbrook2013}, the authors propose
		a controller for single-input-single-output linear time-invariant
		systems to asymptotically track reference trajectories or reject disturbances
		of certain frequency $\omega$. The controller is parameterised by
		the frequency and has the transfer function of 
		\[
		G_{c}\left(s\right)=D\left(s\right)C\left(s\right),
		\]
		where $D\left(s\right)$ a stabilising controller for the closed-loop
		system in the absence of $C\left(s\right)$, and 
		\begin{equation}
		C\left(s\right)=\frac{s^{2}+2\zeta\omega s+\omega^{2}}{s^{2}+\omega^{2}}\label{eq:IMC}
		\end{equation}
		is a notch-like internal model controller that enables asymptotic
		regulation. It is shown in \cite{Esbrook2013} that the closed-loop
		with the inclusion of $C\left(s\right)$ remains stable for sufficiently
		small $\zeta$ and $\omega$. It can be checked that $G_{c}\left(s\right)$
		can be realised in the form of (\ref{eq:xyc}) when $\omega$ is a
		function of $\epsilon$ and/or $N$. \end{exmp}
	
	The following assumption is placed on the AFM system dynamics. For
	the ease of notation, denote $x\triangleq\left(\xi_{o},\xi_{c},\xi_{zo},\xi_{zc}\right)$
	and rewrite (\ref{eq:xyo})\textendash (\ref{eq:zc}) in the $\tau$-timescale
	as follows: 
	\begin{align}
	\frac{dx}{d\tau} & =f\left(x,\xi_{r},\epsilon,N\right),\label{eq:x}\\
	e & =h\left(x,\xi_{r}\right),\label{eq:e}\\
	e_{z} & =h_{z}\left(x\right),\label{eq:ez}
	\end{align}
	where the specific forms of $f$, $h$ and $h_{z}$ are omitted for
	space reasons. 
	
	\begin{ass} \label{ass:steady-state} There exist $\overline{\epsilon},\underbar{\ensuremath{N}}>0$
		such that for each $0<\epsilon\leq\overline{\epsilon}$ and $N\geq\underbar{\ensuremath{N}}$
		and each $\xi_{r}\left(0\right)\in D_{r}$ of (\ref{eq:xyr2}), the
		AFM system dynamics (\ref{eq:x}) has a uniformly globally asymptotically
		stable (UGAS) steady-state solution $\bar{x}\left(\tau\right)$\footnote{Notice the dependence of $\bar{x}\left(\tau\right)$, $\bar{e}\left(\tau\right)$
			and $\bar{e}_{z}\left(\tau\right)$ on $\epsilon$, $N$ and $\xi_{r}\left(0\right)$
			is dropped to ease the notation. } which satisfies that 
		\begin{enumerate}
			\item there exists a sufficiently smooth map $M$ such that 
			\[
			\bar{x}\left(\tau\right)=M\left(\xi_{r}\left(\tau\right),\epsilon,N\right)
			\]
			holds for $\tau\in\mathbb{R}$; 
			\item the steady-state XY tracking error denoted by $\bar{e}\left(\tau\right)\triangleq h\left(\bar{x}\left(\tau\right),\xi_{r}\left(\tau\right)\right)$
			satisfies 
			\[
			\bar{e}\left(\tau\right)=0
			\]
			for $\tau\in\mathbb{R}$; and 
			\item there exists some constant $k_{e}>0$ such that the steady-state Z
			regulation error $\bar{e}_{z}\left(\tau\right)\triangleq h_{z}\left(\bar{x}\left(\tau\right)\right)$
			satisfies 
			\[
			\left|\bar{e}_{z}\left(\tau+T_{l}\right)-\bar{e}_{z}\left(\tau\right)\right|\leq\frac{k_{e}}{N}
			\]
			for $\tau\in\mathbb{R}$. 
		\end{enumerate}
		Moreover, there exist functions $\alpha_{x1},\alpha_{x2}\in\mathcal{K}_{\infty},\alpha_{f}\in\mathcal{K}$
		and a smooth Lyapunov function $V_{x}\left(\tilde{x},\epsilon,N\right)$
		such that for each $0<\epsilon\leq\overline{\epsilon}$, $N\geq\underbar{\ensuremath{N}}$
		and $\xi_{r}\left(0\right)\in D_{r}$, 
		\begin{align*}
		\alpha_{x1}\left(\left|\tilde{x}\right|\right)\leq V_{x}\left(\tilde{x},\epsilon,N\right) & \leq\alpha_{x2}\left(\left|\tilde{x}\right|\right)\\
		\frac{\partial V_{x}}{\partial\tilde{x}}\tilde{f}\left(\tilde{x},\xi_{r}\left(\tau\right),\epsilon,N\right) & \leq-\alpha_{f}\left(\left|\tilde{x}\right|\right)
		\end{align*}
		holds for all $\tau\geq0$, where {\small{}$\tilde{x}\triangleq x-M\left(\xi_{r}\left(\tau\right),\epsilon,N\right)$}
		and {\small{}$\tilde{f}\left(\tilde{x},\xi_{r}\left(\tau\right),\epsilon,N\right)\triangleq f\left(\tilde{x}+M\left(\xi_{r}\left(\tau\right),\epsilon,N\right),\xi_{r}\left(\tau\right),\epsilon,N\right)-f\left(M\left(\xi_{r}\left(\tau\right),\epsilon,N\right),\xi_{r}\left(\tau\right),\epsilon,N\right)$}.
		
	\end{ass} 
	\begin{rem}
		Due to linearity, the XY subsystem (\ref{eq:xyo})\textendash (\ref{eq:xyc})
		and the Z subsystem (\ref{eq:zo})\textendash (\ref{eq:zc}) can be
		stabilised under mild assumptions on the actuator dynamics. For the
		same reason, it can be seen that the stabilised AFM system (\ref{eq:x})
		is \emph{uniformly convergent with the uniformly bounded steady-state
			property} \cite{Pavlov2007} with respect to bounded continuous input
		signals, and the existence of the map $M$ follows from \cite[Theorem 2]{Pavlov2007}. 
	\end{rem}
	\begin{comment}
	Easy to see that the mapping 
	\[
	\xi\left(\tau\right)=M_{xy}\left(\xi_{r}\left(\tau\right),\epsilon,N\right)
	\]
	exists due to linearity. Also easy to see that the Z system 
	\[
	\frac{d}{d\tau}\xi_{z}=\frac{1}{\epsilon}A_{z}\xi_{z}+\frac{1}{\epsilon}B_{z}d_{z}
	\]
	is uniformly convergent with the uniformly bounded steady-state property.
	Therefore, the auxiliary interconnection 
	\begin{align*}
	\frac{d}{d\tau}\xi_{r} & =A_{r}\left(N\right)\xi_{r}\\
	\frac{d}{d\tau}\xi_{z} & =\frac{1}{\epsilon}A_{z}\xi_{z}+\frac{1}{\epsilon}B_{z}f_{s}\circ CM_{xy}\left(\xi_{r}\left(\tau\right),\epsilon,N\right)
	\end{align*}
	has a mapping 
	\[
	\xi_{z}\left(\tau\right)=M_{z}\left(\xi_{r}\left(\tau\right),\epsilon,N\right)
	\]
	Therefore the mapping $M$ could be defined as 
	\[
	M=\begin{bmatrix}M_{xy}\\
	M_{z}
	\end{bmatrix}
	\]
	\end{comment}
	\begin{rem}
		Zero steady-state XY tracking error can be achieved by output regulation
		controllers based on some practical assumptions on the reference trajectory
		model (\ref{eq:xyr}) and the XY actuator dynamics (\ref{eq:xyo}).
		See \cite{Esbrook2013,Yong2014,Bazaei2016} for nanopositioning applications
		and \cite{Francis1976} for linear systems in general. This may not
		be achievable in practice, but this does not obstruct the implementation
		of the proposed scheme in general, as will be shown in the simulation
		result section. %
		\begin{comment}
		However, no research has been conducted on constructing the XY controllers
		for a range of frequencies such that $V_{x}$ exists and satisfied
		the given conditions. 
		\end{comment}
	\end{rem}
	\begin{rem}
		For the aforementioned scan patterns, two consecutive scan lines are
		typically very close in order to achieve a high image resolution.
		This implies that the topography signal, and therefore the Z regulation
		error, would be close to periodic with the period $T_{l}$ in the
		$\tau$-timescale. Moreover, as the number of scan lines increases,
		the differences between two consecutive lines and between the corresponding
		regulation errors would decrease. This observation is the key to the
		proposed adaptive scan. 
	\end{rem}
	\begin{rem}
		\label{rem:uniformity} The requirement on existence of $\alpha_{x1}$,
		$\alpha_{x2}$ and $\alpha_{f}$ being independent of $N$ may be
		difficult to satisfy for arbitrarily large $N$. In practice, however,
		$N$ is upper bounded by the finite resolution of the XY subsystem
		and hence $\alpha_{x1}$, $\alpha_{x2}$ and $\alpha_{f}$ be established
		for the largest possible $N$. 
	\end{rem}
	
	\subsection{Definition of Performance Function}
	
	\begin{comment}
	The proposed adaptive scan varies $\epsilon$ online to maximise some
	performance measure, which is a function of the regulation error and
	scan rate over one scan line. Thus, it is necessary to pinpoint the
	time instant that corresponds to the start of the current scan line,
	which becomes easy in the $\tau$-timescale (see Remark \ref{rem:tau-timescale}).
	\end{comment}
	
	In order to generalise the intuitive performance metric mentioned
	in Section \ref{sec:INTRO}, the performance function is assumed to
	have the form of 
	\begin{align}
	q\left(\tau\right) & =Q\left(e_{z}\left[\tau\right],\epsilon\left[\tau\right]\right)\label{eq:q}\\
	& \triangleq g_{e}\left(L_{e}\left(e_{z}\left[\tau\right]\right)\right)+g_{s}\left(L_{s}\left(\epsilon\left[\tau\right]\right)\right),\nonumber 
	\end{align}
	where $g_{e}$ and $g_{s}$ are arbitrary sufficiently smooth functions
	and $L_{e}$ and $L_{s}$ are defined by 
	\begin{align}
	L_{e}\left(e_{z}\left[\tau\right]\right) & \triangleq\begin{cases}
	{\displaystyle \left(\frac{1}{T_{l}}\int_{\tau-T_{l}}^{\tau}\left|e_{z}\left(\sigma\right)\right|^{p}d\sigma\right)^{\frac{1}{p}}} & p\in\left[1,\infty\right)\\
	{\displaystyle \max_{\sigma\in\left[\tau-T_{l},\tau\right]}\left|e_{z}\left(\sigma\right)\right|} & p=\infty
	\end{cases},\label{eq:qe}\\
	L_{s}\left(\epsilon\left[\tau\right]\right) & \triangleq\epsilon\left(\tau-\tau_{\phi}\right),\label{eq:qs}
	\end{align}
	respectively. Specifically, $L_{e}\left(e_{z}\left[\tau\right]\right)$
	is the $L_{p}$-norm characterisation of the line regulation error,
	and $L_{s}\left(\epsilon\left[\tau\right]\right)$ is the delayed
	scan rate, where $p$ is a design parameter and $\tau_{\phi}\in\left[0,T_l\right]$ is
	a small constant to match the effective time delays between these
	two channels. The functions $g_{e}$ and $g_{s}$ can be chosen to
	suit application-specific imaging goals and requirements, e.g. minimising
	the deviation from some target local imaging accuracy. Mathematically,
	the arguments of $Q$ are histories of regulation error and scan rate:
	$e_{z}\left[\tau\right]\triangleq\left\{ e_{z}\left(\tau+\sigma\right):\sigma\in\left[-\tau_{d},0\right]\right\} $
	and $\epsilon\left[\tau\right]\triangleq\left\{ \epsilon\left(\tau+\sigma\right):\sigma\in\left[-\tau_{d},0\right]\right\} $
	with $\tau_{d}=T_{l}+\frac{2\pi}{\omega}$ being the length of the
	histories, where $\omega$ is defined later. %
	\begin{comment}
	calculation procedure: 
	\begin{align*}
	& =-\left(\frac{1}{T_{l}}\int_{\tau\left(t\right)-T_{l}}^{\tau\left(t\right)}\left|e_{z}\left(t\left(\sigma\right)\right)\right|^{p}d\sigma\right)^{\frac{1}{p}}\\
	& =-\left(\frac{1}{T_{l}}\int_{t\left(\tau\left(t\right)-T_{l}\right)}^{t\left(\tau\left(t\right)\right)}\left|e_{z}\left(t\left(\tau\left(\sigma\right)\right)\right)\right|^{p}\frac{d\tau}{d\sigma}\left(\sigma\right)d\sigma\right)^{\frac{1}{p}}\\
	& =-\left(\frac{1}{T_{l}}\int_{t\left(\tau\left(t\right)-T_{l}\right)}^{t}\left|e_{z}\left(\sigma\right)\right|^{p}\epsilon\left(\sigma\right)d\sigma\right)^{\frac{1}{p}}
	\end{align*}
	\end{comment}
	
	In steady state, the relation between fixed values of $\epsilon,N$
	and $q$ can be defined by the following time-varying map for all
	$\tau\in\mathbb{R}$: 
	\begin{align*}
	& \bar{J}\left(\epsilon,N,\tau\right)\\
	& \triangleq g_{e}\left(\left(\frac{1}{T_{l}}\int_{\tau-T_{l}}^{\tau}\left|h_{z}\left(M\left(\xi_{r}\left(\sigma\right),\epsilon,N\right)\right)\right|^{p}d\sigma\right)^{\frac{1}{p}}\right)+g_{s}\left(\epsilon\right).
	\end{align*}
	
	\begin{ass} \label{ass:maximum} The following holds: 
		\begin{itemize}
			\item $\bar{J}\left(\epsilon,N,\tau\right)$, $\frac{\partial\bar{J}}{\partial\epsilon}\left(\epsilon,N,\tau\right)$
			and $\frac{\partial^{2}\bar{J}}{\partial\epsilon^{2}}\left(\epsilon,N,\tau\right)$
			are bounded for all $0<\epsilon\leq\overline{\epsilon}$ and $N\geq\underbar{\ensuremath{N}}$
			uniformly in $\tau\in\mathbb{R}$. 
			\item $\bar{J}\left(\epsilon,N,\tau\right)$ has a maximum at the solution
			$\epsilon^{*}\left(N,\tau\right)$ for each $N\geq\underbar{\ensuremath{N}}$
			and $\tau\in\mathbb{R}$. 
			\item There exists $\alpha_{J}\in\mathcal{K}$ such that 
			\[
			\frac{\partial\bar{J}}{\partial\epsilon}\left(\epsilon,N,\tau\right)\left(\epsilon-\epsilon^{*}\left(N,\tau\right)\right)\leq-\alpha_{J}\left(\left|\epsilon-\epsilon^{*}\left(N,\tau\right)\right|\right)
			\]
			holds for all $0<\epsilon\leq\overline{\epsilon}$, $N\geq\underbar{\ensuremath{N}}$
			and $\tau\in\mathbb{R}$. 
			\item There exist $\underbar{\ensuremath{\epsilon^{*}}},\overline{\epsilon^{*}}\in\left(0,\overline{\epsilon}\right]$
			such that $\underbar{\ensuremath{\epsilon^{*}}}\leq\ensuremath{\epsilon^{*}}\left(N,\tau\right)\leq\overline{\epsilon^{*}}$
			holds for all $N>\underbar{\ensuremath{N}}$ and $\tau\in\mathbb{R}$. 
		\end{itemize}
	\end{ass}
	\begin{rem}
		The satisfaction of Assumption \ref{ass:maximum} depends on the sample
		topography, the AFM system dynamics and the design of $g_{e}$ and
		$g_{s}$. In general, it may be difficult to analytically verify Assumption
		\ref{ass:maximum}. However, an example is given in Section \ref{sec:SR}
		using a carefully constructed sample. 
	\end{rem}
	\begin{ass} \label{ass:slowly-varying} The functions $\bar{J}\left(\epsilon,N,\tau\right)$
		and $\epsilon^{*}\left(N,\tau\right)$ are continuously differentiable
		in $\tau$. Moreover, there exist $k_{J},k_{\epsilon}>0$ such that
		\begin{align*}
		\left|\frac{\partial\bar{J}}{\partial\tau}\right| & \leq\frac{k_{J}}{N},\\
		\left|\frac{\partial\epsilon^{*}}{\partial\tau}\right| & \leq\frac{k_{\epsilon}}{N},
		\end{align*}
		for all $0<\epsilon\leq\overline{\epsilon}$ and $N\geq\underbar{\ensuremath{N}}$
		uniformly in $\tau\in\mathbb{R}$. \end{ass}
	\begin{rem}
		Assumption \ref{ass:slowly-varying} states that the steady-state
		cost map and optimal scan rate trajectory become less time-dependent
		as the number of scan lines $N$ increases. Indeed, as $N$ approaches
		infinity, the scan is conducted over the same scan line, yielding
		a constant steady-state cost map and optimal scan rate solution. 
	\end{rem}
	
	\section{Adaptation Law Design}
	
	\label{sec:ES}
	
	The performance function defines a unique optimal scan rate trajectory.
	Since the sample topography is unknown, the optimal trajectory is
	also unknown. Therefore, an adaptation law is devised to search for
	the optimal scan rate trajectory based on the extremum-seeking approach.
	Theoretical performance guarantees are then provided by conducting
	stability analysis on the overall system.
	
	\subsection{Extremum-Seeking Controller}
	
	\begin{figure}[!t]
		\centering \includegraphics[width=8.4cm]{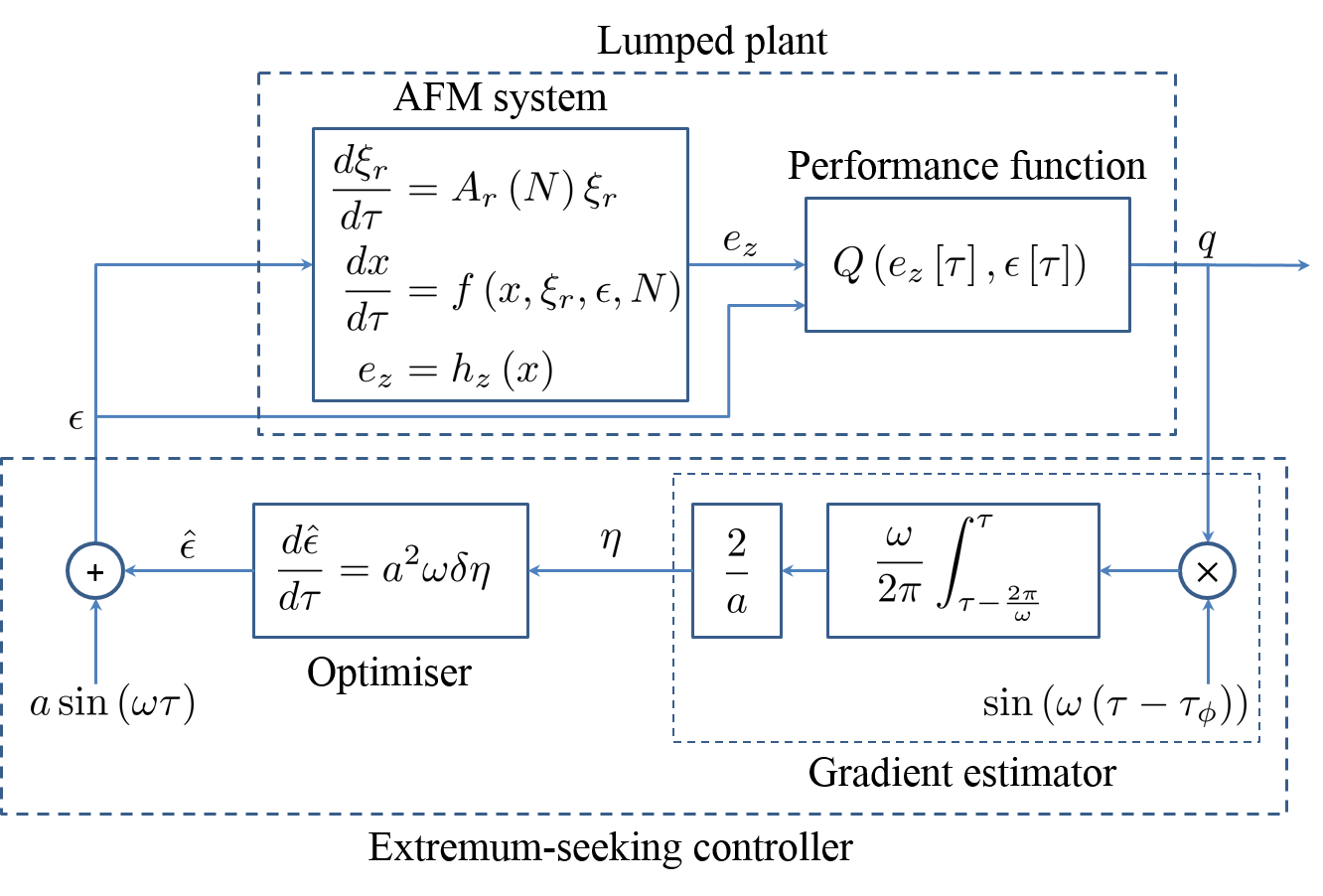}
		\caption{Overall system block diagram. Notice that the plant is represented
			in the $\tau$-timescale, and the performance function, dither, gradient
			estimator and optimiser are all based in the $\tau$-timescale.}
		\label{fig:es_block_diagram} 
	\end{figure}
	
	The closed-loop system is illustrated in Fig. \ref{fig:es_block_diagram}.
	As in \cite{Haring2013}, a dither is added to the estimate of $\epsilon^{*}$,
	$\hat{\epsilon}$, yielding $\epsilon=\hat{\epsilon}+a\sin\left(\omega\tau\right)$,
	where $a$ is the dither amplitude and $\omega$ is the dither frequency.
	Notice the dither is based in the $\tau$-timescale and $\tau$ can
	be obtained by integrating $\epsilon$ with respect to $t$ (see (\ref{eq:tau})).
	
	Then, the resulting performance output from the lumped plant is passed
	through a mean-over-perturbation-period (MOPP) filter that is used
	for gradient estimation: 
	\begin{equation}
	\eta\left(\tau\right)=\frac{\omega}{a\pi}\int_{\tau-\frac{2\pi}{\omega}}^{\tau}q\left(\sigma\right)\sin\left(\omega\left(\sigma-\tau_{\phi}\right)\right)d\sigma,\label{eq:eta}
	\end{equation}
	\begin{comment}
	\[
	=\frac{\epsilon\left(t\right)\omega}{a\pi}\int_{t\left(\tau-\frac{2\pi}{\omega}\right)}^{t}q\left(t\right)\sin\left(\omega\left(\tau\left(\sigma\right)-\tau_{\phi}\right)\right)\epsilon\left(\sigma\right)d\sigma
	\]
	\end{comment}
	where $\eta$ is the estimated gradient of $\bar{J}$ and $\tau_{\phi}$
	is defined earlier.
	Unlike commonly used finite-order low/high-pass filters, the MOPP
	completely filters out the higher-order harmonics of $\omega$, thus
	improving the accuracy of gradient estimation and increasing the convergence
	speed \cite{Haring2013}.
	
	Finally, the optimiser has the form of 
	\begin{equation}
	\frac{d\hat{\epsilon}}{d\tau}=a^{2}\omega\delta\eta,\label{eq:eps_hat}
	\end{equation}
	where $\delta>0$ is a constant parameter.
	\begin{rem}
		Increasing $a$, $\omega$ and $\delta$ can increase the adaptation
		speed, however, large $a$ and $\omega$ result in a larger and faster
		steady-state performance metric fluctuation, respectively. Moreover,
		to ensure stability, $\omega$ needs to be sufficiently smaller than
		$\omega_{l}$ to ensure timescale separation between lumped plant
		dynamics and gradient estimation, and similarly, $\delta$ needs to
		be sufficiently small to ensure timescale separation between gradient
		estimation and scan rate optimisation. A qualitative guideline on
		parameter tunings that is sufficient for the overall system stability
		is given in Theorem \ref{theorem:sgpas}. 
	\end{rem}
	\begin{rem}
		The implementation of the adaptive scan scheme requires the evaluations
		of (\ref{eq:tau}) and (\ref{eq:q})\textendash (\ref{eq:eps_hat}),
		which only involve two numerical integrations, one moving $L_{p}$-norm
		evaluation and one moving average evaluation. Therefore, the implementation
		of the adaptive scan scheme does not pose significant computational
		requirements. 
	\end{rem}
	
	\subsection{Stability Analysis}
	
	Next, the SGPAS result is to be established in the $\tau$ timescale.
	To begin, define $\tilde{x}=x-M\left(\xi_{r},\epsilon,N\right)$ and
	$\tilde{\epsilon}=\hat{\epsilon}-\epsilon^{*}\left(N,\tau\right)$,
	and the overall system (\ref{eq:xyr2}), (\ref{eq:x}), (\ref{eq:ez})\textendash (\ref{eq:eps_hat})
	can be described as follows: 
	\begin{align}
	\frac{d\xi_{r}}{d\tau} & =A_{r}\left(N\right)\xi_{r},\label{eq:cl1}\\
	\frac{d\tilde{x}}{d\tau} & =\tilde{f}\left(\tilde{x},\xi_{r},\epsilon,N\right)-\frac{\partial M}{\partial\epsilon}\left(\frac{d\hat{\epsilon}}{d\tau}+a\omega\cos\left(\omega\tau\right)\right),\label{eq:cl2}\\
	\frac{d\tilde{\epsilon}}{d\tau} & =\frac{a\omega^{2}\delta}{\pi}\int_{\tau-\frac{2\pi}{\omega}}^{\tau}J\left(\tilde{x}\left[\sigma\right],\xi_{r}\left[\sigma\right],\epsilon\left[\sigma\right],N\right)s\left(\sigma\right)d\sigma\nonumber \\
	& \phantom{=}-\frac{\partial\epsilon^{*}}{\partial\tau}\left(N,\tau\right),\label{eq:cl3}
	\end{align}
	\begin{comment}
	calculation procedure: 
	\begin{align*}
	\frac{d\tilde{x}}{d\tau} & =\frac{dx}{d\tau}-\frac{\partial M}{\partial\epsilon}\frac{d\epsilon}{d\tau}-\frac{\partial M}{\partial\xi_{r}}\frac{d\xi_{r}}{d\tau}\\
	& =f_{a}\left(\tilde{x}+M,\epsilon,\xi_{r}\right)-\frac{\partial M}{\partial\epsilon}\left(\frac{d\tilde{\epsilon}}{d\tau}+\frac{d\epsilon^{*}}{d\tau}+a\omega\cos\left(\omega\tau\right)\right)-\frac{\partial M}{\partial\xi_{r}}A_{r}\xi_{r}\\
	& =f_{a}\left(\tilde{x}+M,\epsilon,\xi_{r}\right)-f_{a}\left(M,\epsilon,\xi_{r}\right)-\frac{\partial M}{\partial\epsilon}\left(\frac{d\tilde{\epsilon}}{d\tau}+\frac{d\epsilon^{*}}{d\tau}+a\omega\cos\left(\omega\tau\right)\right)\\
	& \triangleq\tilde{f}_{a}\left(\tilde{x},\epsilon,\xi_{r}\right)-\frac{\partial M}{\partial\epsilon}\left(\frac{d\tilde{\epsilon}}{d\tau}+\frac{d\epsilon^{*}}{d\tau}+a\omega\cos\left(\omega\tau\right)\right)
	\end{align*}
	\begin{align*}
	\frac{d\tilde{\epsilon}}{d\tau} & =\frac{d\hat{\epsilon}}{d\tau}-\frac{d\epsilon^{*}}{d\tau}\\
	& =\frac{a\omega^{2}\delta}{\pi}\int_{\tau-\frac{2\pi}{\omega}}^{\tau}J\left(\tilde{x}\left[\tau\right],\epsilon\left[\tau\right],\xi_{r}\left[\tau\right]\right)\sin\left(\omega\left(\sigma-\tau_{\phi}\right)\right)d\sigma-\frac{d\epsilon^{*}}{d\tau}
	\end{align*}
	\end{comment}
	where $\tilde{f}\left(\tilde{x},\xi_{r},\epsilon,N\right)$ is defined
	in Assumption \ref{ass:steady-state}, $J\left(\tilde{x}\left[\tau\right],\xi_{r}\left[\tau\right],\epsilon\left[\tau\right],N\right)\triangleq Q\left(h_{z}\left(\tilde{x}\left[\tau\right]+M\left(\xi_{r}\left[\tau\right],\epsilon\left[\tau\right],N\right)\right),\epsilon\left[\tau\right]\right)$
	with $\tilde{x}\left[\tau\right]$ and $\xi_{r}\left[\tau\right]$
	defined in the same way as $\epsilon\left[\tau\right]$, and $s\left(\tau\right)\triangleq\sin\left(\omega\left(\tau-\tau_{\phi}\right)\right)$.
	
	\begin{theorem} \label{theorem:sgpas} Suppose Assumptions \ref{ass:reference}\textendash \ref{ass:slowly-varying}
		hold, then for each $\rho_{x}^{0},\rho_{\epsilon}^{0}>0$, each $\nu_{\epsilon}^{+}\in\left(0,\frac{\overline{\epsilon}}{2}\right)$
		and each $\nu_{x},\nu_{\epsilon}>0$, there exist $\rho_{x},\rho_{\epsilon}>0$
		and $a^{*}>0$ such that for each $a\in\left(0,a^{*}\right)$ there
		exists $\omega^{*}>0$ such that for each $\omega\in\left(0,\omega^{*}\right)$
		there exists $\delta^{*}>0$ such that for each $\delta\in\left(0,\delta^{*}\right)$
		such that there exists $N^{*}>0$ such that for each $N\in\left(N^{*},\infty\right)$
		and each initial condition $\left(\tilde{x}\left[0\right],\tilde{\epsilon}\left[0\right]\right)$
		such that $\begin{bmatrix}\left|\tilde{x}\left[0\right]\right|\\
		\left|\tilde{\epsilon}\left[0\right]\right|
		\end{bmatrix}\leq\begin{bmatrix}\rho_{x}^{0}\\
		\rho_{\epsilon}^{0}
		\end{bmatrix}$ and $\nu_{\epsilon}^{+}\leq\hat{\epsilon}\left(0\right)\leq\overline{\epsilon}-\nu_{\epsilon}^{+}$,
		the solution of (\ref{eq:cl1})\textendash (\ref{eq:cl3}) satisfies
		\[
		\sup_{\tau\geq0}\begin{bmatrix}\left|\tilde{x}\left(\tau\right)\right|\\
		\left|\tilde{\epsilon}\left(\tau\right)\right|
		\end{bmatrix}\leq\begin{bmatrix}\rho_{x}\\
		\rho_{\epsilon}
		\end{bmatrix}\text{ and }\limsup_{\tau\rightarrow\infty}\begin{bmatrix}\left|\tilde{x}\left(\tau\right)\right|\\
		\left|\tilde{\epsilon}\left(\tau\right)\right|
		\end{bmatrix}\leq\begin{bmatrix}\nu_{x}\\
		\nu_{\epsilon}
		\end{bmatrix}
		\]
		and $0<\epsilon\left(\tau\right)<\overline{\epsilon}$ for all $\tau\geq0$.
		
	\end{theorem} 
	\begin{IEEEproof}
		See Appendix. 
	\end{IEEEproof}
	\begin{cor}
		\label{cor:sgpas} Under the conditions of Theorem \ref{theorem:sgpas},
		the solution of (\ref{eq:cl1})\textendash (\ref{eq:cl3}) satisfies
		\[
		\sup_{t\geq0}\begin{bmatrix}\left|\tilde{x}\left(t\right)\right|\\
		\left|\tilde{\epsilon}\left(t\right)\right|
		\end{bmatrix}\leq\begin{bmatrix}\rho_{x}\\
		\rho_{\epsilon}
		\end{bmatrix}\text{ and }\limsup_{t\rightarrow\infty}\begin{bmatrix}\left|\tilde{x}\left(t\right)\right|\\
		\left|\tilde{\epsilon}\left(t\right)\right|
		\end{bmatrix}\leq\begin{bmatrix}\nu_{x}\\
		\nu_{\epsilon}
		\end{bmatrix}.
		\]
	\end{cor}
	\begin{rem}
		Theorem \ref{theorem:sgpas} together with Corollary \ref{cor:sgpas}
		states the SGPAS of the overall system and signifies that the XY tracking
		errors converge to a neighbourhood of zero and the estimate of optimal
		scan rate converges to a neighbourhood of the true value provided
		parameters $a$, $\omega$, $\delta$ and $N$ are properly tuned.
		Specifically, for an arbitrarily small given neighbourhood, a sufficiently
		small $a$ is firstly selected, based on which then a sufficiently
		small $\omega$ is selected, based on which a sufficiently small $\delta$
		is selected, and finally based on which a sufficiently large $N$
		is selected. These selections of parameters can then ensure the convergence
		to such a small neighbourhood. 
	\end{rem}

	\begin{figure*}[b]
		\hrulefill{} 
		\begin{align}
		G_{xo}\left(s\right) & =\frac{245s^{2}-2.73\times10^{6}s+2.9\times10^{10}}{s^{3}+1796s^{2}+3.39\times10^{7}s+5.64\times10^{10}},\label{eq:tf1}\\
		G_{yo}\left(s\right) & =\frac{241s^{4}-3.18\times10^{6}s^{3}+4.26\times10^{10}s^{2}-9.41\times10^{13}s+1.13\times10^{18}}{s^{5}+2328s^{4}+6.79\times10^{7}s^{3}+1.45\times10^{11}s^{2}+1.14\times10^{15}s+2.21\times10^{18}},\\
		G_{zo}\left(s\right) & =\frac{64.6s^{4}-2.77\times10^{5}s^{3}+5.27\times10^{11}s^{2}-3.41\times10^{16}s+1.5\times10^{21}}{s^{5}+4.52\times10^{4}s^{4}+3.64\times10^{9}s^{3}+1.14\times10^{11}s^{2}+3\times10^{18}s+3.16\times10^{22}}.\label{eq:tf3}
		\end{align}
	\end{figure*}

	\begin{rem}
		As noted in Remark \ref{rem:uniformity}, there is a finite upper
		bound to the number of scan lines, $N$, in any practical system.
		This limits the explicit implementation of the theorem. However, in
		practice this is not a significant imposition for real systems as
		will be seen in the following result sections. 
	\end{rem}

	\begin{rem}
		Although the initial condition $\hat{\epsilon}\left(0\right)$ cannot
		be selected arbitrarily on $ \mathbb{R} $ due to realistic constraints
		on the scan rate, the set of permissible initial conditions can become
		arbitrarily close to the set of permissible scan rate of the original
		system as specified in Assumption \ref{ass:steady-state}, hence the term `semi-global'. 
	\end{rem}

\begin{figure}[H]
\begin{minipage}[t]{0.5\columnwidth}%
	\includegraphics[width=4.2cm]{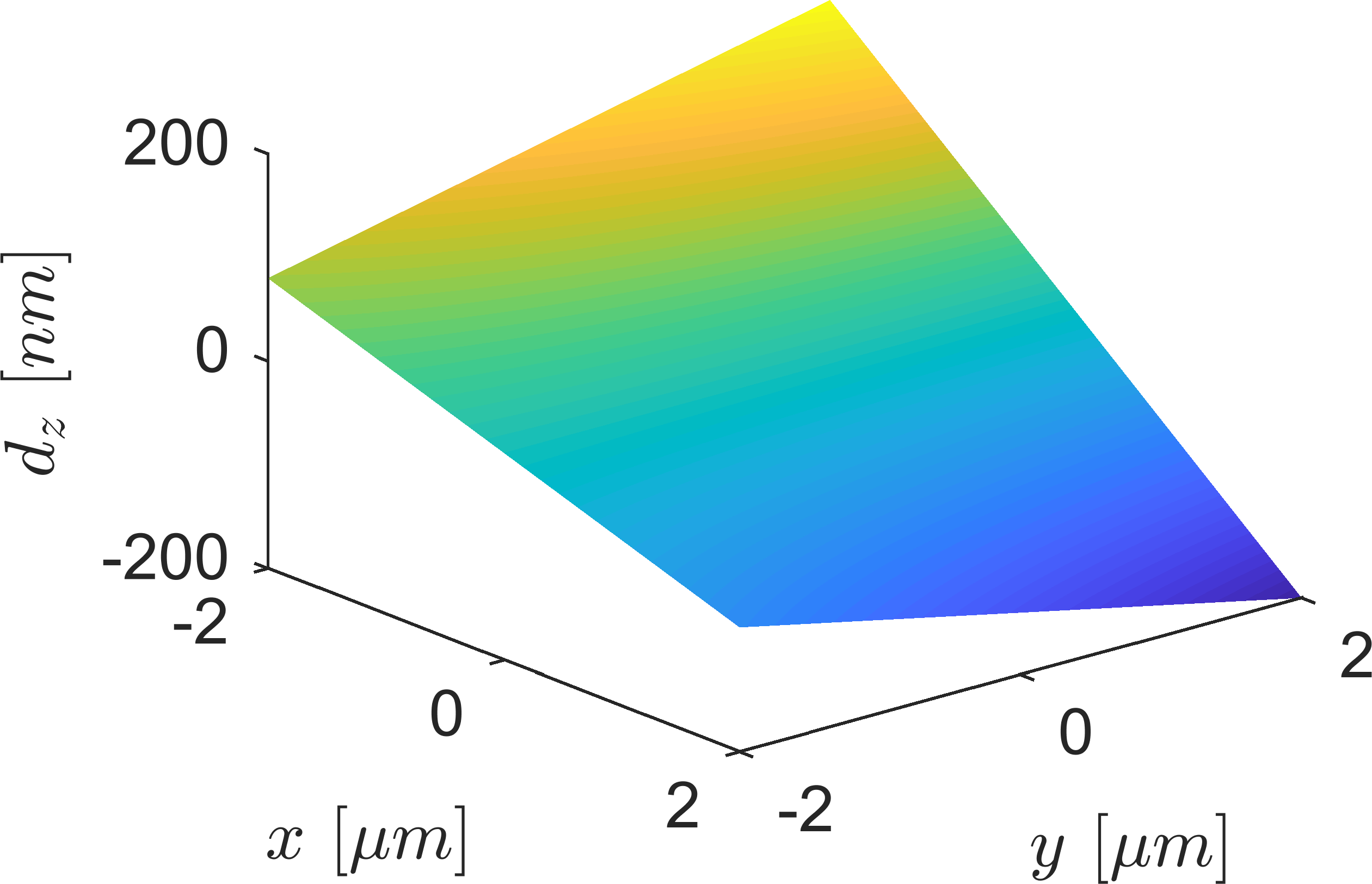} %
\end{minipage}\hfill{}%
\begin{minipage}[t]{0.5\columnwidth}%
	\includegraphics[width=4.2cm]{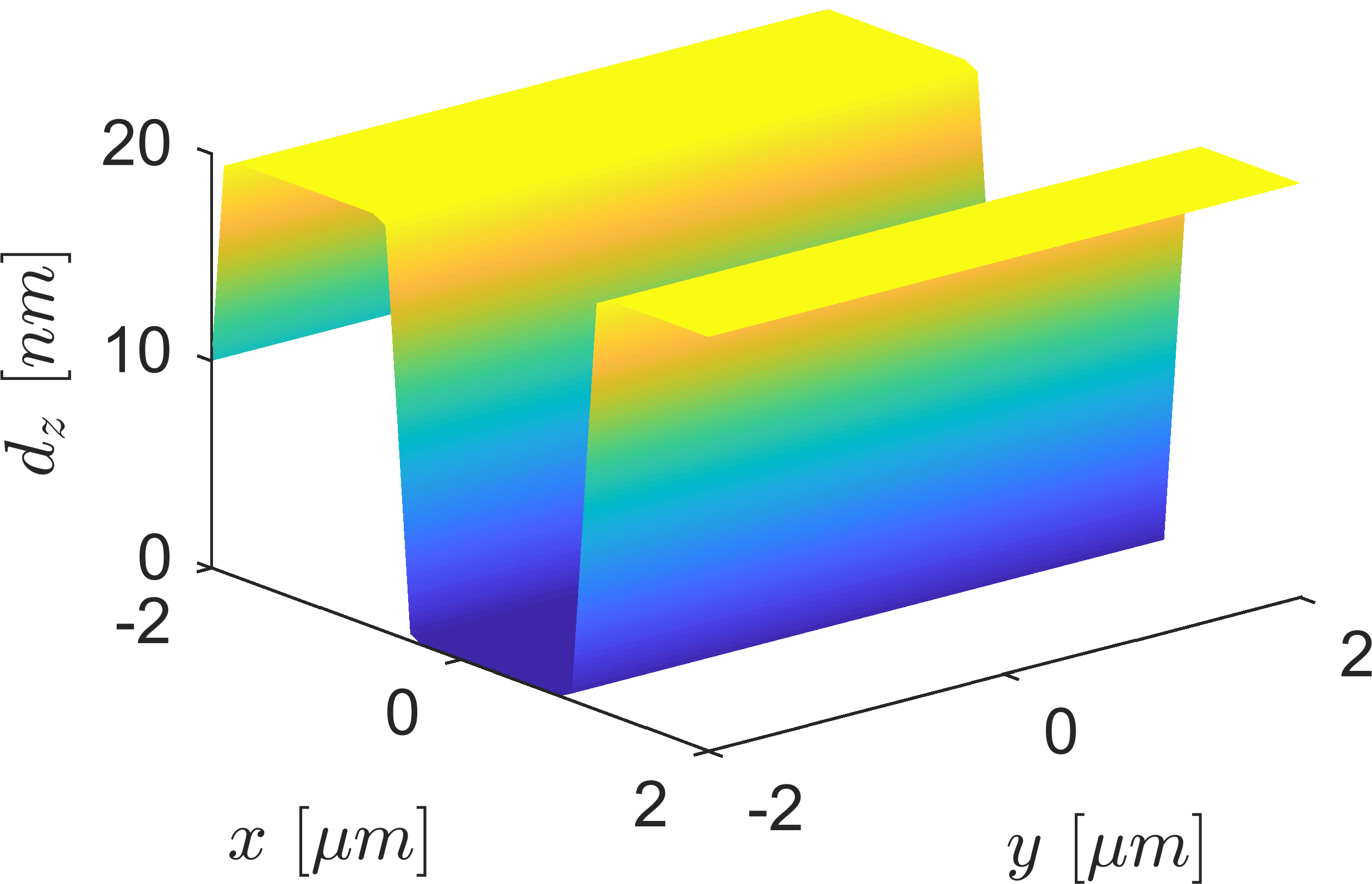} %
\end{minipage}\caption{Sample topography. Scenario One: a tilted and bent plane (left). Scenario
	Two: periodic rectangular features (right).}
\label{fig:topography} 
\end{figure}
	
	\section{Simulation Results}
	
	\label{sec:SR}
	
	\begin{comment}
	The structure can follow Esbrook2013An. 
	\end{comment}
	
	Simulation studies are conducted based on two scenarios to verify
	the theoretical results as well as to provide motivating scenarios
	under which the adaptive scan is desirable. Specifically, Scenario
	One provides a thorough exposition of the theoretical results by studying
	a constructed sample, and Scenario Two suggests a solution to the
	known issue of varying linear velocity and resulting imaging accuracy
	during a spiral scan. For both scenarios, the linear models (\ref{eq:tf1})\textendash (\ref{eq:tf3})
	are used for the X, Y and Z actuators, respectively.
	
	\subsection{Scenario One: Constructed Sample and Sinusoidal Raster Scan}
	
	In this scenario, the sample topography map (Fig. \ref{fig:topography},
	left) represents a tilted and bent plane whose magnitude of the gradient
	along X increases along Y. The map has the expression of $f_{s}\left(x,y\right)=-\left(15y+70\right)x$.
	The sinusoidal raster scan in Example \ref{exmp:sra} is conducted
	with a scan size of $L=\SI{4}{\micro\meter}$ and a nominal line period
	of $T_{l}=\SI{0.01}{\second}$. This suggests a nominal scan duration
	of $T_{0}=\SI{0.995}{s}$ if $N=100$. It can be shown that Assumptions
	\ref{ass:reference} and \ref{ass:topography} are satisfied.
	
	The XY controller design follows Example \ref{exmp:3}. The stabilising
	controllers for X and Y axes are both $D\left(s\right)=\frac{667}{s}$.
	The X internal model controller is designed with $\zeta_{xc}=1$ and
	$\omega_{xc}=\epsilon\omega_{l}$. A slight modification is made for
	the Y internal model controller design to enhance convergence, resulting
	the controller %
	\begin{comment}
	The initial design of the Y internal model controller has a similar
	form but turns out to result in an unacceptably slow convergence.
	After inspection, it is found that this is because the internal model
	controller has a pair of slow zeros placed nearby the slow poles,
	which generates a pair of root loci in the vicinity and constrains
	the speed of corresponding closed-loop poles. A tweak of the design
	is made by moving the slow zeros further to the left-hand plane, and
	the final controller has the form of 
	\end{comment}
	$C_{y}\left(s\right)=\frac{\left(s+500\left(\epsilon\omega_{fr}\right)\right)^{2}}{s^{2}+\left(\epsilon\omega_{fr}\right)^{2}}$.
	For the Z axis, an integral controller is used with a gain of $7.14\times10^{3}$.
	It can be checked that the controllers can be realised in the form
	of (\ref{eq:xyc}) and (\ref{eq:zc}) and Assumption \ref{ass:steady-state}
	holds for any compact subset of $N\in\left[\underbar{\ensuremath{N}},\infty\right)$.
	
	\begin{comment}
	The controller design yields gain margins of more than $16$ dB and
	$19$ dB and phase margins of more than $99^{\circ}$ and $30^{\circ}$
	for X and Y axes respectively for all $0<\epsilon\leq2$ and $50\leq N\leq400$.
	The settling time for tracking a step is less than $20$ ms and $28$
	ms in the $\tau$-timescale for X and Y axes respectively for all
	$\epsilon,N$ in the given range. For the Z axis, an integral controller
	is used with an integral gain of $10^{4}$, which yields a gain margin
	of $9.66$ dB, a phase margin of $89.8^{\circ}$ and a settling time
	less than $0.8$ ms in the $\tau$-timescale for all $\epsilon,N$
	in the range. This suggests that Assumption \ref{ass:steady-state}
	is satisfied, but only for $0<\epsilon\leq2$ and $50\leq N\leq400$.
	However, the stability result should not be affected as long as $N$
	is chosen to be no larger than $400$. 
	\end{comment}
	
	The following performance function is used with $p=\infty$ chosen
	in (\ref{eq:qe}): 
	\[
	Q\left(e_{z}\left[\tau\right],\epsilon\right)=-\left(\max_{\sigma\in\left[\tau-T_{l},\tau\right]}\left|e_{z}\left(\sigma\right)\right|-e_{z}^{*}\right)^{2}
	\]
	with $e_{z}^{*}=\SI{4}{\nano\meter}$. This represents the objective
	of keeping the maximum regulation error for each line at \SI{4}{\nano\meter}.
	Notice that since the movement along Y is much slower than that along
	X, the steady-state topography signal could be locally well approximated
	by a sinusoidal signal whose frequency is the actual scan rate $\omega_l\epsilon$
	and whose amplitude depends on the local gradient of the sample along
	X, i.e. 
	\begin{align*}
	A\left(N,\tau\right) & =A_{l}\left|\left.\frac{\partial f_{s}}{\partial x}\right|_{y=A_{l}\sin\left(\omega_{fr}\tau-\frac{\pi}{2}\right)}\right|\\
	& =\left(15A_{l}\sin\left(\omega_{fr}\tau-\frac{\pi}{2}\right)+70\right)A_{l}.
	\end{align*}
	Moreover, since the Z closed-loop dynamics are linear, denote the
	sensitivity function by $S_{z}\left(s\right)$, then the maximum line
	regulation error can be approximated by $A\left(N,\tau\right)S_{z}\left(\omega_l\epsilon\right)$.
	Finally, since $S_{z}\left(s\right)$ is almost linear in the passband,
	it may be approximated by a proportional function with an identified
	gain of $k_{S_{z}}=2.95\times10^{-3}$. Therefore, the approximate
	steady-state cost map is 
	\[
	\hat{\bar{J}}\left(\epsilon,N,\tau\right)=-\left(A\left(N,\tau\right)k_{S_{z}}\omega_l\epsilon-e_{z}^{*}\right)^{2}
	\]
	and the approximate optimal scan rate trajectory is 
	\[
	\hat{\epsilon}^{*}\left(N,\tau\right)=\frac{e_{z}^{*}}{A\left(N,\tau\right)k_{S_{z}}\omega_l}.
	\]
	It can be verified that Assumptions \ref{ass:maximum} and \ref{ass:slowly-varying}
	are satisfied.

\begin{figure}[!t]
		\centering \includegraphics[width=8.4cm]{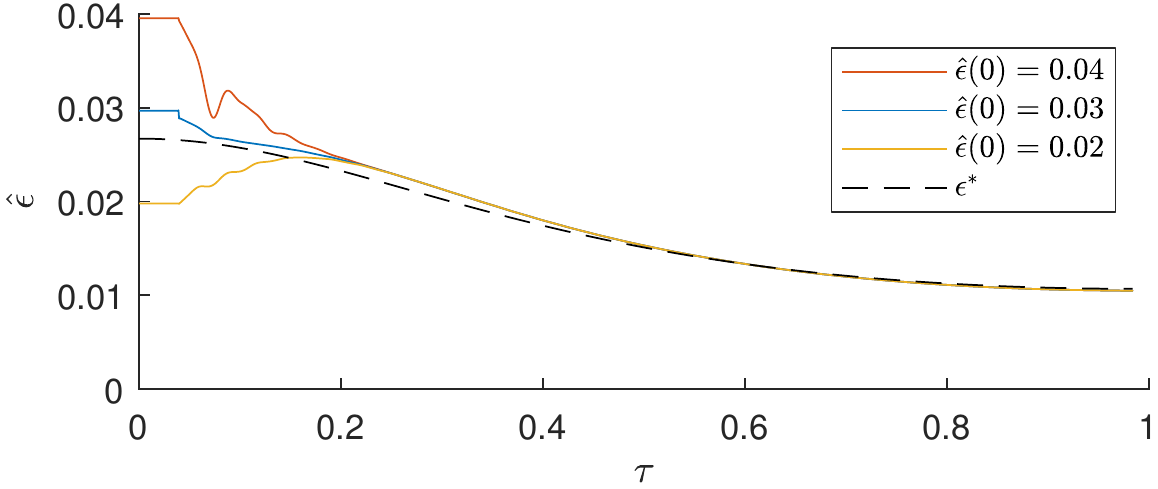}
		\caption{Convergence of scan rate from different initial conditions under Scenario
			One.}
		\label{fig:sim1_1}
		
		\vspace{0.7cm}
		
		\centering \includegraphics[width=8.4cm]{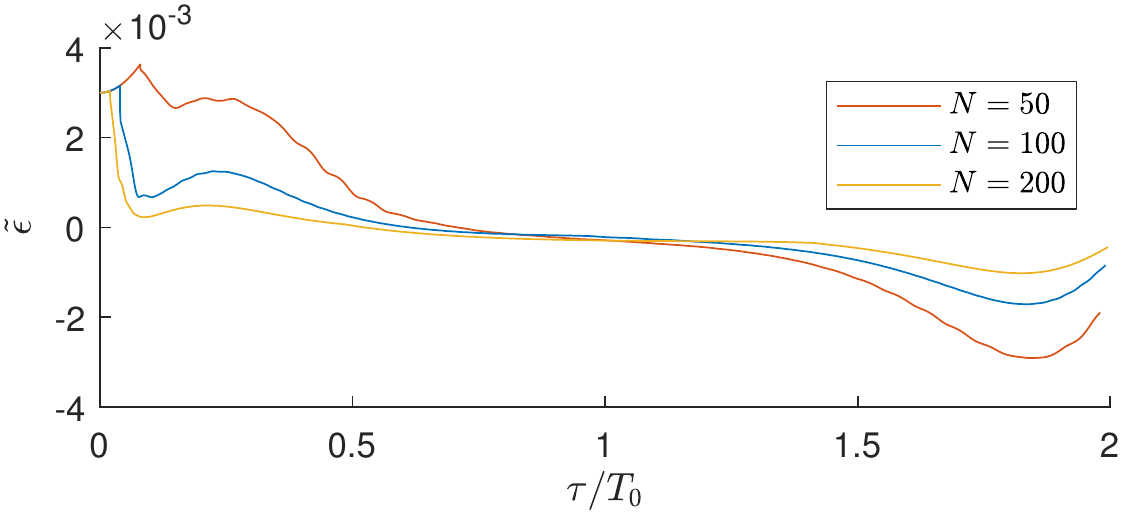}
		\caption{Adaptation error reduces as $N$ increases under Scenario One.}
		\label{fig:sim1_2}
		
		\vspace{0.7cm}
		
		\centering \includegraphics[width=8.4cm]{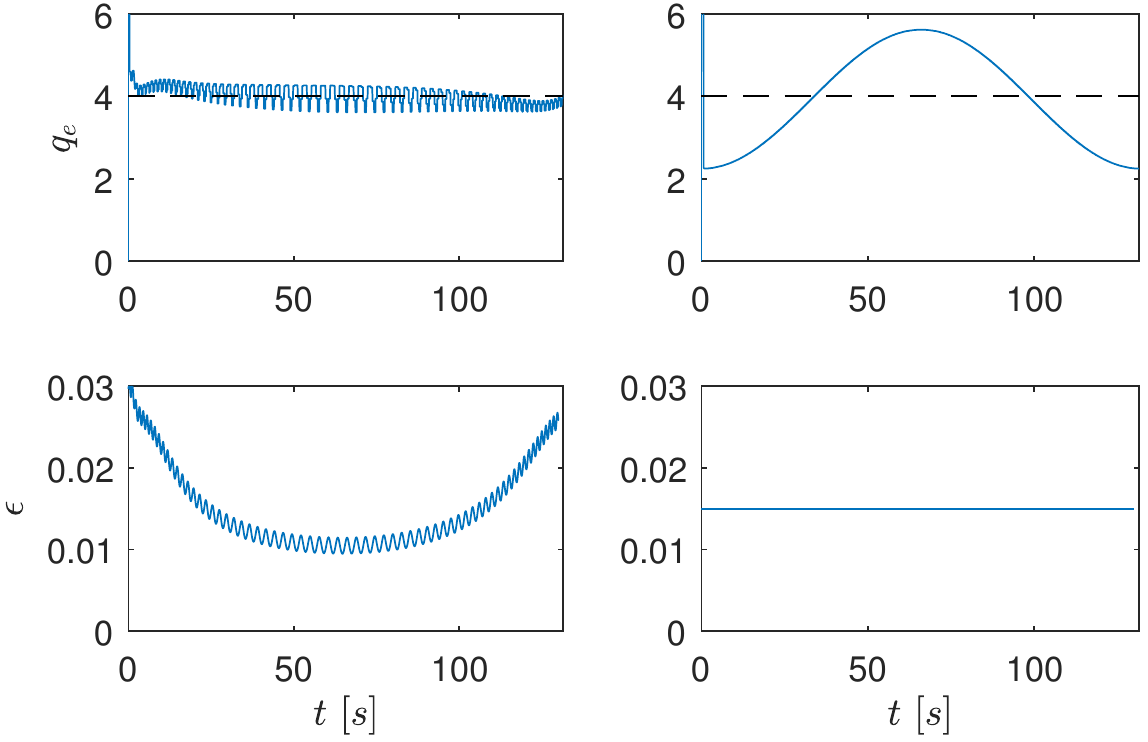}
		\caption{Comparison between adaptive (left) and fixed-rate scan (right) under
			Scenario One.}
		\label{fig:sim1_4} 
	\end{figure}

	With Assumptions \ref{ass:reference}\textendash \ref{ass:slowly-varying}
	now satisfied, it is expected that the SGPAS result from Theorem \ref{theorem:sgpas}
	holds for the closed-loop system. This is confirmed by the simulation
	results shown in Figs. \ref{fig:sim1_1} and \ref{fig:sim1_2} with
	the ES parameters $\tau_{\phi}=6\times10^{-3}$, $a=0.001$, $\omega=210$
	and $\delta=2$. In Fig. \ref{fig:sim1_1}, $N$ is set to $100$
	and three sets of scans are conducted with different initial conditions
	$\hat{\epsilon}\left(0\right)$. It can be seen that all three trajectories
	of $\hat{\epsilon}\left(\tau\right)$ converge to the neighbourhood
	of the approximate optimal solution $\hat{\epsilon}^{*}\left(N,\tau\right)$.
	Note that in order to improve the transient behaviour, the optimiser
	is not turned on until $\tau=T_{l}+\frac{2\pi}{\omega}$. It may also
	be worth noting that a uniform convergence speed is attained in the
	$\tau$-timescale rather than in the $t$-timescale.
	
	Fig. \ref{fig:sim1_2} shows the adaptation error trajectory for two
	scans (forward and backward along Y) under different $N$ values.
	It can be seen that the adaptation error reduces as $N$ increases.
	Note that the X axis is normalised and represents the number of completed
	scans.
	
	Finally, the performance of the adaptive scan method when $N=100$
	is compared with that of the conventional fixed-rate scan method.
	This is done by setting the scan rate of the fixed-rate method such
	that it would complete the scan under the same duration. The XY reference
	trajectories, XY and Z controllers remain unchanged for the fixed-rate
	case. The result is shown in Fig. \ref{fig:sim1_4}. It can be seen
	that, despite sample topography variations, the proposed method adjusts
	the scan rate to track the desired imaging accuracy, i.e. the maximum
	regulation error over scan lines, $q_{e}$, is maintained at about
	the set point $e_{z}^{*}=4$ \si{nm}. The small oscillations of
	$q_{e}$ and $\epsilon$ are due to the action of dithering. On the
	other hand, the fixed-rate method yields a considerable contrast in
	imaging accuracy in different regions of the sample.
	
	\subsection{Comparison with Existing Adaptive Scan under Conventional Raster
		Scan}
	
	In this subsection, the proposed adaptive scan scheme is compared
	with the existing adaptive scan method in \cite{Ren2016} using the
	same sample. Since the latter only considers conventional raster scan
	pattern, the following triangular XY reference trajectories are used:
	\begin{align*}
	r_{x}\left(\tau\right) & =A_{l}\tri\left(\omega_{l}\tau-\frac{\pi}{2}\right),\\
	r_{y}\left(\tau\right) & =A_{l}\tri\left(\omega_{fr}\tau-\frac{\pi}{2}\right),
	\end{align*}
	where 
	\[
	\tri\left(\tau\right)\triangleq\begin{cases}
	\frac{2}{\pi}\tau & \tau-2k\pi\in\left[0,\frac{\pi}{2}\right)\\
	2-\frac{2}{\pi}\tau & \tau-2k\pi\in\left[\frac{\pi}{2},\frac{3\pi}{2}\right)\\
	-4+\frac{2}{\pi}\tau & \tau-2k\pi\in\left[\frac{3\pi}{2},2\pi\right)
	\end{cases},\ k\in\mathbb{Z}.
	\]
	
\begin{figure}[!t]
		\centering \includegraphics[width=8.4cm]{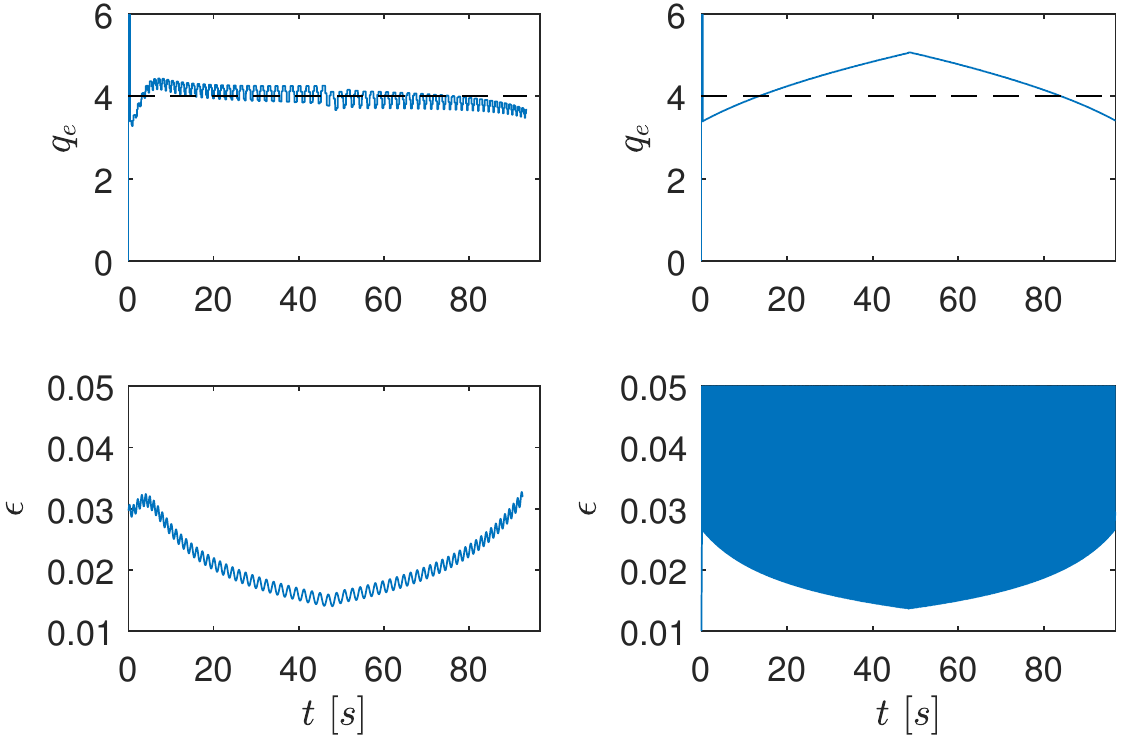}
		\caption{Comparison between the proposed adaptive scan and the existing adaptive
			scan \cite{Ren2016}.}
		\label{fig:sim3_1}
		
		\vspace{0.7cm}
		
		\centering \includegraphics[width=8.4cm]{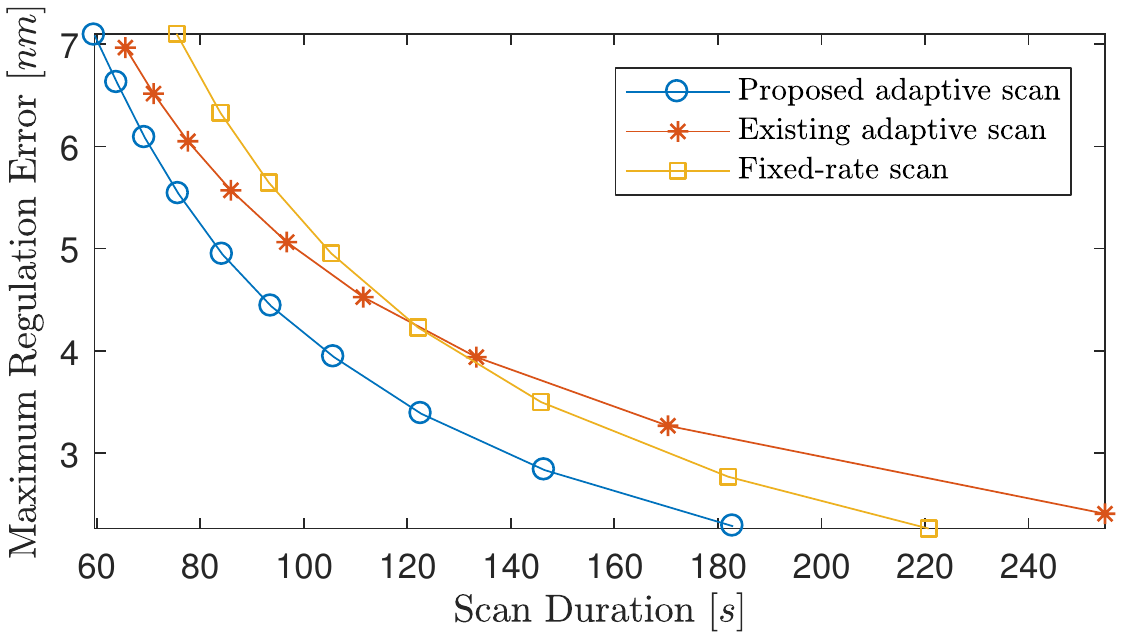}
		\caption{Scan duration and maximum steady-state regulation error achieved by
			the proposed adaptive scan, existing adaptive scan \cite{Ren2016},
			and fixed-rate scan. }
		\label{fig:sim3_2} 
	\end{figure}

	For the proposed adaptive scan, the XY and Z controllers, performance
	function, and ES parameters remain unchanged. Since the XY reference
	trajectories are non-smooth, asymptotic tracking is not possible,
	and Assumptions \ref{ass:reference} and \ref{ass:steady-state} no
	longer hold. However, as shown in the left column of Fig. \ref{fig:sim3_1},
	the adaptive scan still manages to maintain $q_{e}$ at about the
	set point $e_{z}^{*}$ by adapting the scan rate.
	
	The adaptive scan proposed in \cite{Ren2016} is then implemented
	for comparison. The same XY and Z controllers are used. It should
	be noted that \cite{Ren2016} combines a few field-specific insights
	into its control strategy, but only the scan rate adaptation component
	is replicated here for the comparison. Although it is not clear how
	the sample topography gradient necessary for the adaptation law is
	obtained in \cite{Ren2016}, the true sample gradient is known for
	Scenario One and is directly fed to the adaptation law. The parameters
	are chosen to be $\kappa=1$, $\nabla h^{*}=50$ and $d^{*}=3$. The
	desired traversing speed $\dot{x}^{*}$ is chosen to be \SI{10}{\micro\meter\per\second}
	to achieve a similar scan duration for a fair comparison.
	
	The result of the existing adaptation law is shown in the right column
	of Fig. \ref{fig:sim3_1}. It can be seen that the adaptation law
	does not maintain the moving maximum of regulation error at the specified
	value, despite it taking slightly longer to finish the scan. Also,
	in contrast to the proposed adaptation law, it reacts immediately
	to regulation error changes, resulting in large fluctuations in the
	scan rate (bottom right of Fig. \ref{fig:sim3_1}). This may impose
	heavy burdens on the XY actuators and controllers and worsen the XY
	tracking performance.
	
	To more systematically compare the two adaptation laws, a series of
	simulations are conducted by varying $e_{z}^{*}$ and $\dot{x}^{*}$,
	respectively. The steady-state maximum regulation error and scan duration
	are recorded for each simulation. The fixed-rate scan method is also
	conducted for benchmarking. The result is shown in Fig. \ref{fig:sim3_2}.
	It can be seen that the proposed adaptation law outperforms both fixed-rate
	scan method and the existing adaptation law. Furthermore, the existing
	adaptation law does not yield consistently better results than the
	fixed-rate scan. This is perhaps because such an ad-hoc algorithm
	does not necessarily lead to optimality.
	
	\begin{figure}[!t]
		\centering \includegraphics[width=8.4cm]{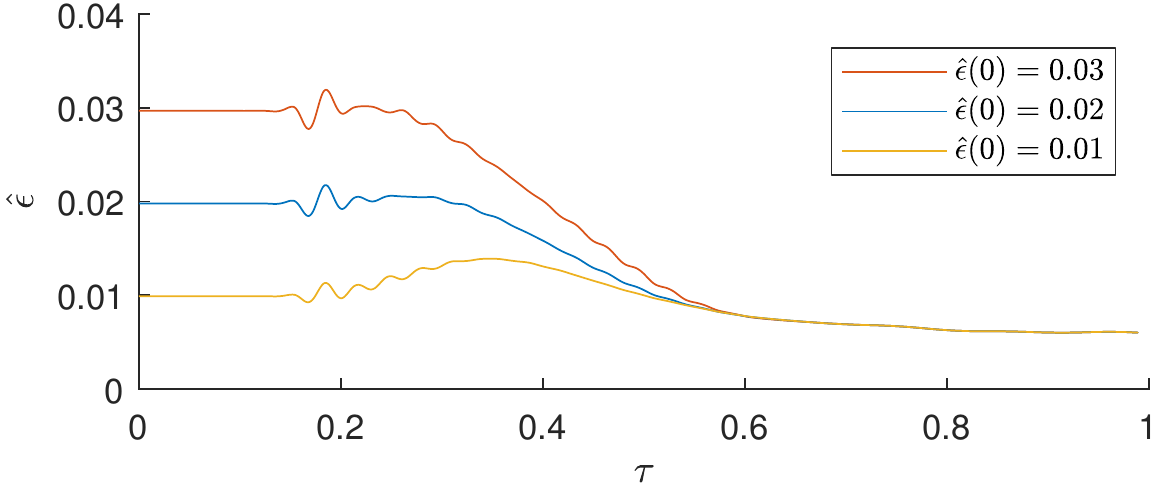}
		\caption{Convergence of scan rate from different initial conditions under Scenario
			Two in simulation.}
		\label{fig:sim2_1}
		
		\vspace{0.55cm}
		
		\centering \includegraphics[width=8.4cm]{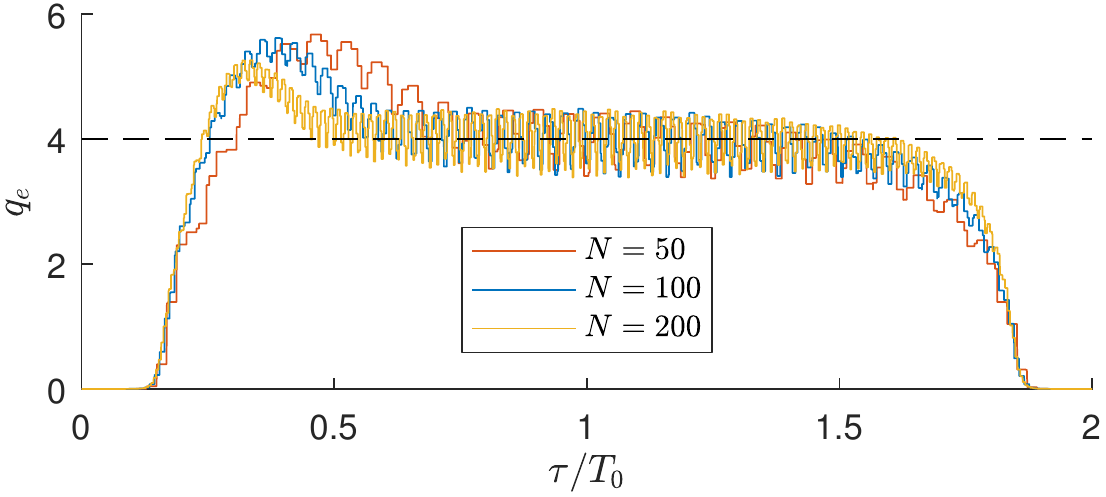}
		\caption{Adaptation error reduces as $N$ increases under Scenario Two in simulation.}
		\label{fig:sim2_2}
		
		\vspace{0.55cm}
		
		\centering \includegraphics[width=8.4cm]{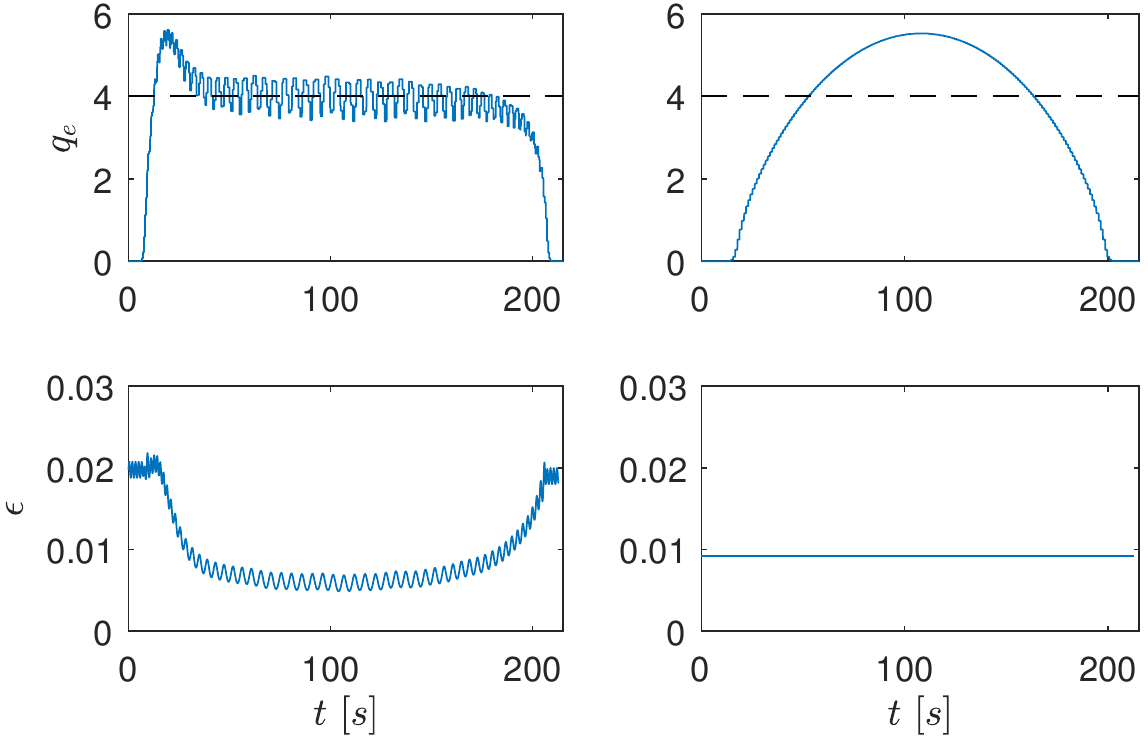}
		\caption{Comparison between adaptive (left) and fixed-rate scan (right) under
			Scenario Two in simulation.}
		\label{fig:sim2_4} 
	\end{figure}
	
	\begin{figure}[!t]
		\centering \includegraphics[width=8.4cm]{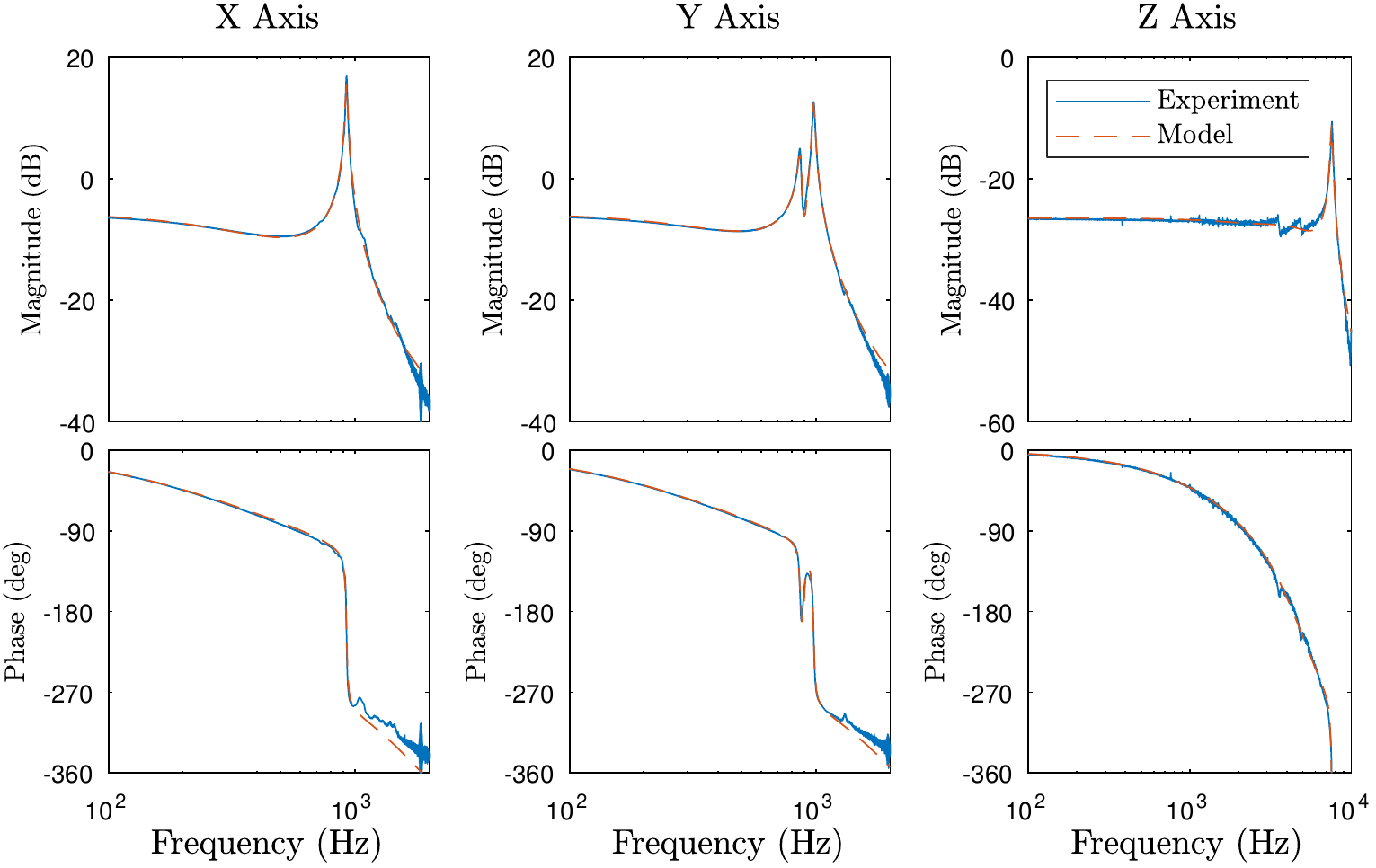}\caption{System identification result.}
		\label{fig:sys_id} 
	\end{figure}
	
	\begin{figure}[!t]
		\centering \includegraphics[width=8.4cm]{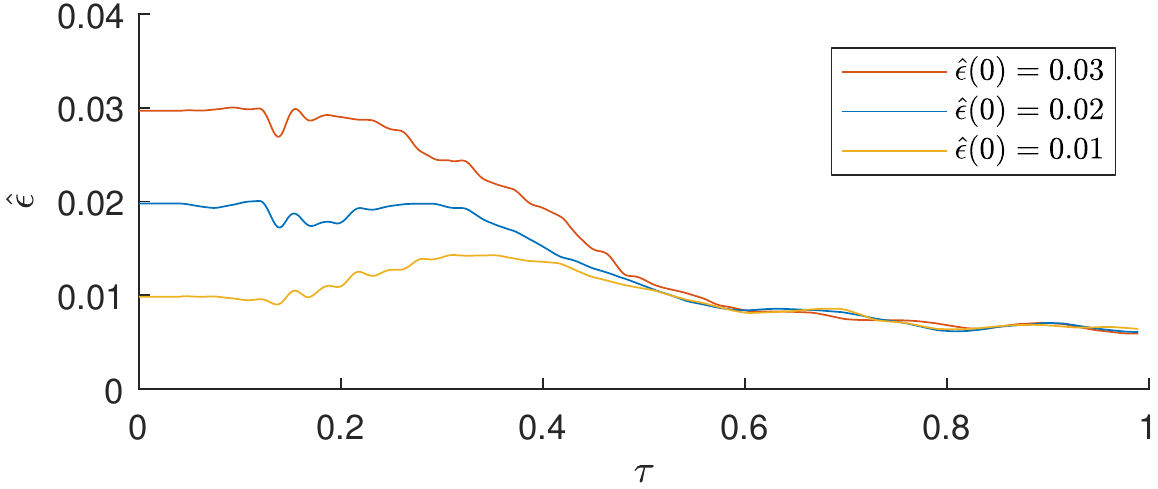}
		\caption{Convergence of scan rate from different initial conditions under Scenario
			Two in experiment.}
		\label{fig:exp2_1}
		
		\vspace{0.5cm}
		
		\centering \includegraphics[width=8.4cm]{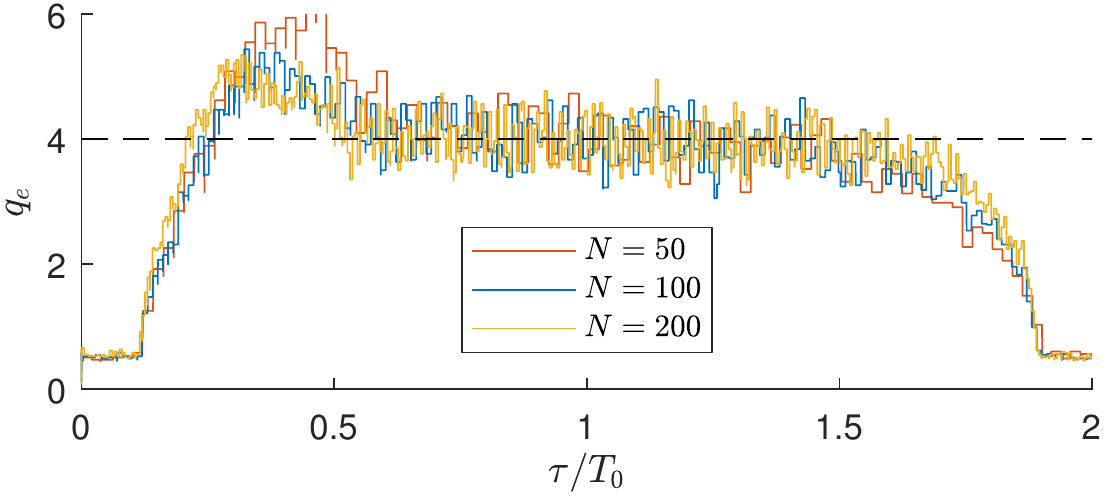}
		\caption{Adaptation error reduces as $N$ increases under Scenario Two in experiment.}
		\label{fig:exp2_2}
		
		\vspace{0.5cm}
		
		\centering \includegraphics[width=8.4cm]{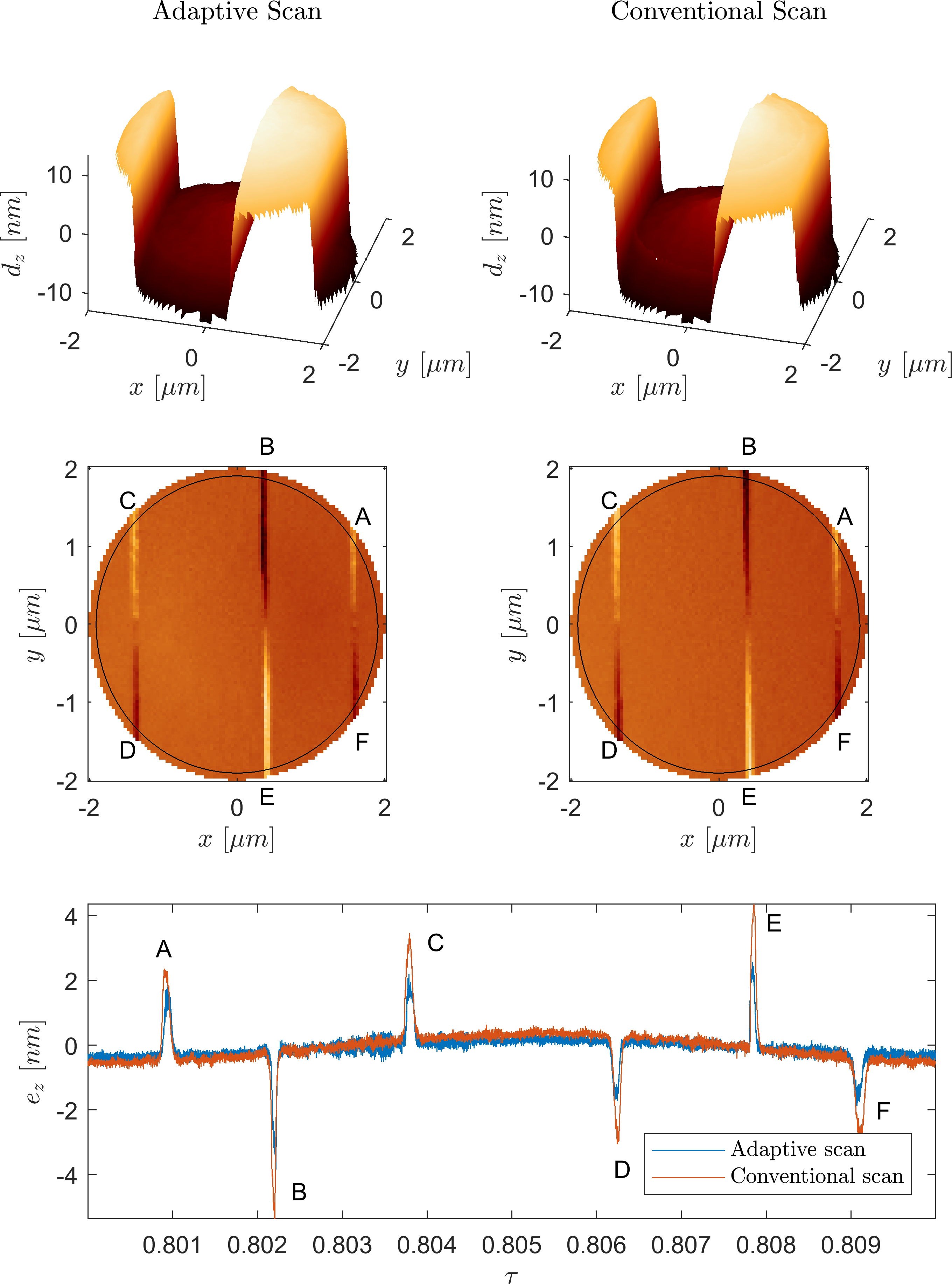}
		\caption{Comparison between adaptive (left) and fixed-rate scan (right) under
			Scenario Two in experiment. From top to bottom: topography image,
			deflection image (regulation error), and error trajectory along one
			scan line indicated in the deflection image. The letters draw connections
			between features as shown in the deflection image and in the regulation
			error trajectory. }
		\label{fig:exp2_7} 
	\end{figure}
	
	\subsection{Scenario Two: Periodic Rectangular Features and Spiral Scan}
	
	In order to demonstrate the generality of the proposed approach, another
	simulation study is conducted using a different scan pattern and different
	sample topography. In this scenario, a virtual NT-MDT TGZ1 calibration
	grating with periodic rectangular features of heights $h=20$ \si{nm}
	is modelled with the superposition of 2D sigmoid functions (Fig. \ref{fig:topography},
	right). Their shapes are tuned such that the generated regulation
	error are the same as that measured with the real TGZ1 sample. The
	spiral scan pattern described in Example \ref{exmp:spi} is applied
	with $L=\SI{4}{\micro\meter}$ and $T_{l}=\SI{0.01}{\second}$, which
	implies a nominal scan duration of $T_{0}=\SI{1}{s}$ for a $100$-line
	scan.
	
	The XY controller design is similar to the previous case except for
	the internal model. In particular, the stabilising controllers for
	both X and Y axes remain $D\left(s\right)=\frac{667}{s}$, and for
	each of the axes two internal modes are used in the form of Example
	\ref{exmp:3} with $\omega_{1}=\epsilon\left(\omega_{l}+\omega_{fs}\right)$,
	$\omega_{2}=\epsilon\left(\omega_{l}-\omega_{fs}\right)$ and $\zeta_{1}=\zeta_{2}=1$.
	The Z controller and the performance function remain the same. The
	ES parameters are also the same except now $\omega=200$, $\delta=0.5$.
	The optimiser remains inactive until $\tau=T_{l}+\frac{2\pi}{\omega}$.
	
	It should be noted that in this scenario it is unlikely to obtain
	a closed-form expression of the steady-state cost map $\bar{J}$ and
	optimal scan rate trajectory. However, it is known that the spiral
	scan generates a varying linear velocity which impacts on the bandwidth
	of the topography signal and hence on the imaging accuracy \cite{Teo2016a}.
	Therefore, it is expected that the scan rate would adapt accordingly
	to compensate for the variation in linear velocity. The simulation
	results confirm this. In Fig. \ref{fig:sim2_1}, trajectories originating
	from three different initial conditions all converge to the same trajectory
	whose scan rate reduces as the scan progresses and spiral radius becomes
	larger. This is because the linear velocity in the centre is smaller
	than that at the edge of the scan. It may also be worth noting that
	the optimiser seems to be inactive for some period of time. This is
	due to the completely flat region at the centre of the scan where
	the gradient estimator fails to obtain a nonzero gradient.
	
	Next, the effect of $N$ on the adaptation is tested. Since the closed-form
	expression of the optimal scan trajectory is unavailable, instead
	of depicting the $\tilde{\epsilon}$ trajectory, Fig. \ref{fig:sim2_2}
	shows the trajectory of the maximum regulation error over the past
	scan line, $q_{e}\left(\tau\right)\triangleq\max_{\sigma\in\left[\tau-T_{l},\tau\right]}\left|e_{z}\left(\sigma\right)\right|$.
	Three tests are performed with different choices of $N$, and each
	test consists of two consecutive scans (radially outward and inward).
	It can be seen that while $q_{e}$ is kept around the set-point value
	in all three cases, the transient property improves as $N$ increases.
	
	Lastly, the adaptive scan method is again compared with the fixed-rate
	method. As shown in Fig. \ref{fig:sim2_4}, it can be seen that except
	for some initial transient, the adaptive method manages to keep the
	regulation error at about the set-point value, whereas the fixed-rate
	method yields a varying error profile.
	
	\section{Experiment}
	
	\label{sec:ER}
	
	\subsection{Experimental Setup}
	
	The proposed extremum-seeking controller along with the XY trajectory
	generator and AFM image acquisition were implemented on a digital
	signal processing system (dSPACE MicroLabBox). The MicroLabBox is
	programmed using the Real-Time Interface for Simulink at a sample
	frequency of \SI{10}{\kilo\hertz} and software ControlDesk
	is used for the real-time application control. Higher sampling rates
	are possible, but not necessary due to the low XY closed-loop bandwidth
	(approximately \SI{54}{\hertz}). The analogue outputs of the
	MicroLabBox for commanding the XY nanopositioner of the AFM are amplified
	using a high-voltage amplifier (Nanonis HVA4) and connected to the
	signal access module of the AFM (NT-MDT NTEGRA) to directly control
	the internal piezoelectric tube actuator. Inbuilt capacitive sensors
	are used to measure the actuator displacements in X, Y and Z directions.
	
	\subsection{System Identification}
	
	System identification is performed by applying a low amplitude periodic
	chirp signal to the high-voltage amplifier and measuring the capacitive
	sensor outputs with complex averaging. Three different frequency ranges
	are measured to improve the signal-to-noise ratio for low frequencies.
	The high-voltage amplifier and capacitive sensors have a bandwidth
	of approximately \SI{10}{\kilo\hertz}. The measured frequency
	response functions (FRFs) along with the fitted models using subspace
	identification \cite{McKelvey2004} are shown in Fig. \ref{fig:sys_id}.
	An internal time-delay in the NT-MDT AFM system is responsible for
	the phase roll-off. The three fitted transfer functions from high-voltage
	amplifier input to capacitive sensor output are (\ref{eq:tf1})\textendash (\ref{eq:tf3}).
	\begin{comment}
	\begin{align*}
	G_{xo}\left(s\right) & =\frac{244.5s^{2}-2.725\times10^{6}s+2.898\times10^{10}}{s^{3}+1796s^{2}+3.394\times10^{7}s+5.643\times10^{10}}\\
	G_{yo}\left(s\right) & =\frac{240.8s^{4}-3.177\times10^{6}s^{3}+4.264\times10^{10}s^{2}-9.409\times10^{13}s+1.125\times10^{18}}{s^{5}+2328s^{4}+6.794\times10^{7}s^{3}+1.452\times10^{11}s^{2}+1.137\times10^{15}s+2.211\times10^{18}}\\
	G_{zo}\left(s\right) & =\frac{64.56s^{4}-2.768\times10^{5}s^{3}+5.274\times10^{11}s^{2}-3.414\times10^{16}s+1.5\times10^{21}}{s^{5}+4.516\times10^{4}s^{4}+3.637\times10^{9}s^{3}+1.144\times10^{11}s^{2}+3.002\times10^{18}s+3.164\times10^{22}}
	\end{align*}
	\end{comment}
	The bode plot for identified models are compared against the FRFs
	in Fig. \ref{fig:sys_id}. It can be seen that the identified plant
	dynamics well approximate the actual ones.
	
	\subsection{AFM Imaging}
	
	An NT-MDT TGZ1 calibration grating with periodic rectangular features
	of heights $h=20\pm1.5$ \si{nm} is imaged in contact mode using
	a spiral trajectory with a scan range of $\SI{4}{\micro\meter}$ (i.e.
	identical to Scenario Two in Section \ref{sec:SR}). The XY controller,
	performance function and ES parameters are designed as per Scenario
	Two and implemented in dSPACE. The inbuilt Z controller is used which
	generates a similar response to the one in simulation.
	
	Overall, the results closely resemble those in simulation, although
	the responses are noisier due to ambient disturbances, sample contaminations
	and sensor measurement noise. In particular, Fig. \ref{fig:exp2_1}
	demonstrates the convergence to a steady state trajectory from various
	initial conditions and the convergence rates are very similar to those
	found in simulation (Fig. \ref{fig:sim2_1}). Fig. \ref{fig:exp2_2}
	demonstrates that the desired imaging accuracy of \SI{4}{\nano\meter}
	is achieved and the adaptation has better transient characteristics
	with a larger $N$. One may notice that the initial $q_{e}$ is nonzero
	in contrast to the simulation. This is due to the nonzero Z sensor
	noise in the experiment. 
	
	Finally, Fig. \ref{fig:exp2_7} shows the comparison between AFM images for the adaptive and fixed-rate methods under $N=100$ and the same duration. Although not shown here, the regulation error trajectories and scan rate trajectories are very close to the simulation results in Fig. \ref{fig:sim2_4}. 
	It can be seen that both methods generate similar topography images.
	In terms of the deflection image, which is a reflection of the regulation
	error signal, the adaptive method yields a more consistent magnitude
	of deflection, whereas that for the fixed-rate method increases along
	the radius. This is confirmed by the bottom figure which depicts the
	regulation error over the scan line indicated in the deflection image.
	Six regulation error peaks induced by the step features along the
	scan line are indicated by letters A\textendash E. It can be seen
	that the adaptive method reduces the regulation error approximately
	by half towards the edge of the scan owing to the scan rate adaptation.
	
	\begin{comment}
	43\% reduction in regulation error is observed at point D 
	\end{comment}
	
	\section{Conclusion}
	
	\label{sec:CON}
	
	An adaptive scan scheme based on the extremum-seeking control framework
	is proposed for AFM imaging. Under the proposed scheme, the scan rate
	is adapted online to achieve the best imaging performance in terms
	of imaging duration and accuracy as specified by the user. Under a
	few practical assumptions, it is shown that the scan rate converges
	to a neighbourhood of the optimal solution for a variety of scan patterns
	and performance measures. Example applications are provided for which
	the proposed scheme can achieve a more desirable imaging performance
	than the fixed-rate method.
	
	\begin{comment}
	
	\section*{Acknowledgement}
	
	The experimental research was performed at The University of Newcastle,
	Callaghan, NSW, Australia. 
	\end{comment}

	\appendix[Proof of Theorem \ref{theorem:sgpas}]
	
	The proof resembles that of \cite{Haring2013}. However, a notable
	difference is that the proof here deals with a time-varying cost function.
	First, the extremum-seeking scheme (\ref{eq:cl2})\textendash (\ref{eq:cl3})
	is seen as a feedback interconnection between the $\tilde{x}$-dynamics
	and the $\tilde{\epsilon}$-dynamics, and bounds on the solutions
	$\tilde{x}\left(\tau\right)$ and $\tilde{\epsilon}\left(\tau\right)$
	are derived. To do this, Propositions 9\textendash 11 in \cite{Haring2013}
	are replicated here as Propositions 1\textendash 3 with due modifications.
	Finally, the obtained bounds are exploited to prove the main theorem
	using a small-gain argument. To start with, Propositions 1\textendash 3
	are stated below and proofs are given.
	
	\begin{pro} Suppose Assumptions \ref{ass:reference}\textendash \ref{ass:slowly-varying}
		hold, then there exists a function $\beta_{x}\in\mathcal{KL}$ and
		for each $\rho_{x},\rho_{\epsilon}>0$ there exists a function $\gamma_{xC}\in\mathcal{K}$
		such that for all $\sup_{\tau\geq0}\left|\tilde{x}\left[\tau\right]\right|\leq\rho_{x}$,
		$\sup_{\tau\geq0}\left|\tilde{\epsilon}\left[\tau\right]\right|\leq\rho_{\epsilon}$
		and for all bounded values of $a,\omega,\delta>0$ and $N\geq\underbar{\ensuremath{N}}$
		the following bound holds for all $\tau\geq0$: 
		\begin{equation}
		\left|\tilde{x}\left(\tau\right)\right|\leq\max\left\{ \beta_{x}\left(\left|\tilde{x}\left(0\right)\right|,\tau\right),\gamma_{xC}\left(a\omega\right)\right\} .\label{eq:prop1}
		\end{equation}
	\end{pro}
	
	\begin{pro} Suppose Assumptions \ref{ass:reference}\textendash \ref{ass:slowly-varying}
		hold, then there exists a function $\beta_{\epsilon}\in\mathcal{KL}$
		and for each $\rho_{x},\rho_{\epsilon}>0$ there exist functions $\gamma_{\epsilon x},\gamma_{\epsilon C}\in\mathcal{K}$
		such that for all $\sup_{\tau\geq0}\left|\tilde{x}\left[\tau\right]\right|\leq\rho_{x}$,
		$\sup_{\tau\geq0}\left|\tilde{\epsilon}\left[\tau\right]\right|\leq\rho_{\epsilon}$
		and for all bounded values of $a,\omega,\delta>0$ and $N\geq\underbar{\ensuremath{N}}$
		the following bound holds for any $\tau^{*}\geq\tau_{d}$ and all
		$\tau\geq\tau^{*}$: 
		\begin{align}
		\left|\tilde{\epsilon}\left(\tau\right)\right| & \leq\max\left\{ \beta_{\epsilon}\left(\left|\tilde{\epsilon}\left(\tau^{*}\right)\right|,a^{2}\omega\delta\left(\tau-\tau^{*}\right)\right),\vphantom{\left(\sup_{\tau\geq\tau^{*}}\right)}\right.\label{eq:prop2}\\
		& \left.\gamma_{\epsilon x}\left(\frac{1}{a}\sup_{\tau\geq\tau^{*}}\left|\tilde{x}\left[\tau\right]\right|\right),\gamma_{\epsilon C}\left(a+\omega+\delta+{\frac{1}{a^{2}\omega\delta N}}\right)\right\} .\nonumber 
		\end{align}
	\end{pro}
	
	\begin{pro} Suppose Assumptions \ref{ass:reference}\textendash \ref{ass:slowly-varying}
		hold, then for each $\rho_{x},\rho_{\epsilon}>0$ there exists a constant
		$c_{\epsilon}>0$ such that for all $\sup_{\tau\geq0}\left|\tilde{x}\left[\tau\right]\right|\leq\rho_{x}$,
		$\sup_{\tau\geq0}\left|\tilde{\epsilon}\left[\tau\right]\right|\leq\rho_{\epsilon}$
		and for all bounded values of $a,\omega,\delta>0$ and $N\geq\underbar{\ensuremath{N}}$
		the following bound holds for all $\tau\geq0$: 
		\begin{equation}
		\left|\tilde{\epsilon}\left(\tau\right)\right|\leq\left|\tilde{\epsilon}\left(0\right)\right|+\left(a\omega\delta+{\frac{1}{N}}\right)c_{\epsilon}\tau.\label{eq:prop3}
		\end{equation}
	\end{pro}
	
	\subsection{Proof of Proposition 1}
	
	Using $V_{x}$ in Assumption \ref{ass:steady-state} as a Lyapunov
	function candidate for (19) yields {\small{}{} 
		\[
		\frac{dV_{x}}{d\tau}\leq-\alpha_{f}\left(\left|\tilde{x}\right|\right)-\left(\frac{\partial V_{x}}{\partial\tilde{x}}\frac{\partial M}{\partial\epsilon}+\frac{\partial V_{x}}{\partial\epsilon}\right)\left(\frac{d\hat{\epsilon}}{d\tau}+a\omega\cos\left(\omega\tau\right)\right).
		\]
	}Recall that $\hat{\epsilon}\left(\tau\right)=\tilde{\epsilon}\left(\tau\right)+\epsilon^{*}\left(N,\tau\right)$
	and 
	\[
	\frac{d\hat{\epsilon}}{d\tau}=\frac{a\omega^{2}\delta}{\pi}\int_{\tau-\frac{2\pi}{\omega}}^{\tau}J\left(\tilde{x}\left[\sigma\right],\xi_{r}\left[\sigma\right],\epsilon\left[\sigma\right],N\right)s\left(\sigma\right)d\sigma.
	\]
	It can be seen that, given bounded values of $a,\omega,\delta,\frac{1}{N}>0$
	and uniformly bounded trajectories of $\tilde{x}\left(\tau\right),\tilde{\epsilon}\left(\tau\right),\epsilon^{*}\left(N,\tau\right)$
	for all $\tau\geq-\tau_{d}$, there exists some constant $c_{J1}>0$
	such that 
	\[
	\left|J\left(\tilde{x}\left[\sigma\right],\xi_{r}\left[\sigma\right],\epsilon\left[\sigma\right],N\right)s\left(\sigma\right)\right|\leq c_{J1}
	\]
	for all $\sigma\geq-\frac{2\pi}{\omega}$ and therefore 
	\[
	\left|\frac{d\hat{\epsilon}}{d\tau}\right|\leq2a\omega\delta c_{J1}
	\]
	for $\tau\geq0$. Thus, it can be seen that there exists some $C_{x}>0$
	such that 
	\[
	\frac{dV_{x}}{d\tau}\leq-\alpha_{f}\left(\left|\tilde{x}\right|\right)+a\omega C_{x}.
	\]
	Then it follows that 
	\[
	\frac{dV_{x}}{d\tau}\leq-\frac{1}{2}\alpha_{f}\left(\left|\tilde{x}\right|\right)
	\]
	when 
	\[
	\left|\tilde{x}\right|\geq\alpha_{f}^{-1}\left(2a\omega C_{x}\right).
	\]
	Therefore, using Theorem 4.18 in \cite{Khalil2002}, it can be shown
	that Proposition 1 holds with $\gamma_{xC}\left(\sigma\right)\triangleq\alpha_{x1}^{-1}\circ\alpha_{x2}\circ\alpha_{f}^{-1}\left(2C_{x}\sigma\right)$.
	
	\subsection{Proof of Proposition 2}
	
	First rewrite (\ref{eq:cl3}) as 
	\begin{align*}
	\frac{d\tilde{\epsilon}}{d\tau} & =a^{2}\omega\delta\frac{\partial\bar{J}}{\partial\epsilon}\left(\hat{\epsilon}\left(\tau\right),N,{\tau}\right)+a\omega\delta\Delta_{1}+a\omega\delta\Delta_{2}\\
	& \phantom{=}+a\omega\delta\Delta_{3}+a\omega\delta\Delta_{4}-{\frac{\partial\epsilon^{*}}{\partial\tau}\left(N,\tau\right)}
	\end{align*}
	with 
	\begin{align*}
	\Delta_{1} & \triangleq\frac{\omega}{\pi}\int_{\tau-\frac{2\pi}{\omega}}^{\tau}\left(\bar{J}\left(\hat{\epsilon}\left(\sigma\right)+as\left(\sigma\right),N,\sigma\right)\right.\\
	& \phantom{\triangleq}\left.-\bar{J}\left(\hat{\epsilon}\left(\tau\right)+as\left(\sigma\right)\right),N,\tau\right)s\left(\sigma\right)d\sigma,
	\end{align*}
	\begin{align*}
	\Delta_{2} & \triangleq\frac{\omega}{\pi}\int_{\tau-\frac{2\pi}{\omega}}^{\tau}\left(J\left(0,\xi_{r}\left[\sigma\right],\epsilon\left[\sigma\right],N\right)\right.\\
	& \phantom{\triangleq}\left.-\bar{J}\left(\hat{\epsilon}\left(\sigma\right)+as\left(\sigma\right),N,\sigma\right)\right)s\left(\sigma\right)d\sigma,
	\end{align*}
	\begin{align*}
	\Delta_{3} & \triangleq\frac{\omega}{\pi}\int_{\tau-\frac{2\pi}{\omega}}^{\tau}\left(J\left(\tilde{x}\left[\sigma\right],\xi_{r}\left[\sigma\right],\epsilon\left[\sigma\right],N\right)\right.\\
	& \phantom{\triangleq}\left.-J\left(0,\xi_{r}\left[\sigma\right],\epsilon\left[\sigma\right],N\right)\right)s\left(\sigma\right)d\sigma,
	\end{align*}
	\[
	\Delta_{4}\triangleq\frac{\omega}{\pi}\int_{\tau-\frac{2\pi}{\omega}}^{\tau}R\left(\tau,\sigma\right)s\left(\sigma\right)d\sigma,
	\]
	where $R\left(\tau,\sigma\right)$ is the remainder term after Taylor
	expansion which has the expression of $R\left(\tau,\sigma\right)\triangleq a^{2}s^{2}\left(\sigma\right)\int_{0}^{1}\left(1-r\right)$$\frac{\partial^{2}\bar{J}}{\partial\epsilon^{2}}\left(\hat{\epsilon}\left(\tau\right)+as\left(\sigma\right)r,N,\tau\right)dr$.
	
	For $\left|\Delta_{1}\right|$, notice that 
	\begin{align*}
	& \bar{J}\left(\hat{\epsilon}\left(\sigma\right)+as\left(\sigma\right),N,\sigma\right)-\bar{J}\left(\hat{\epsilon}\left(\tau\right)+as\left(\sigma\right),N,\tau\right)\\
	& =\int_{\tau}^{\sigma}\left(\frac{\partial\bar{J}}{\partial\epsilon}\left(\hat{\epsilon}\left(r\right)+as\left(\sigma\right),N,r\right)\frac{d\hat{\epsilon}}{d\tau}\left(r\right)\right.\\
	& \phantom{=}\left.+{\frac{\partial\bar{J}}{\partial\tau}\left(\hat{\epsilon}\left(r\right)+as\left(\sigma\right),N,r\right)}\right)dr
	\end{align*}
	for all $\sigma\in\left[\tau-\frac{2\pi}{\omega},\tau\right]$ and
	all $\tau\geq\frac{2\pi}{\omega}$. %
	\begin{comment}
	Private remark: $\tilde{x}\left(\rho\right),\tilde{\epsilon}\left(\rho\right)$
	defined on $\rho\geq-\tau_{d}=-T_{l}-\frac{2\pi}{\omega}$, so $\int_{r-\frac{2\pi}{\omega}}^{r}J\left(\tilde{x}\left[\rho\right],\epsilon\left[\rho\right],\xi_{r}\left[\rho\right]\right)s\left(\rho\right)d\rho$
	defined on $r\geq0$, so the whole thing is defined on $\sigma,\tau\geq0$,
	so we need $\tau\geq\frac{2\pi}{\omega}$. 
	\end{comment}
	Using a result from the Proof of Proposition 1, it can be seen that
	there exists some constant $c_{J2}>0$ such that 
	\[
	\left|\frac{d\hat{\epsilon}}{d\tau}\left(r\right)\right|\leq a\omega\delta c_{J2}
	\]
	for $r\geq0$. Moreover, by Assumption \ref{ass:slowly-varying},
	\[
	\left|{\frac{\partial\bar{J}}{\partial\tau}\left(\hat{\epsilon}\left(r\right)+as\left(\sigma\right),N,r\right)}\right|\leq\frac{k_{J}}{N}
	\]
	for $r\geq0$. Therefore, it can be seen that there exists some constant
	$C_{\Delta_{1}}>0$ such that 
	\[
	\left|\Delta_{1}\right|\leq\left(a\delta+{\frac{1}{\omega N}}\right)C_{\Delta_{1}}
	\]
	for all $\tau\geq\frac{2\pi}{\omega}$. %
	\begin{comment}
	Question: How to define $\epsilon^{*}\left[\tau\right]$ for $\tau\in\left[-\tau_{d},0\right]$?
	The question is whether $\bar{J}$ can be defined on $\tau\in\left[-\tau_{d},0\right]$. 
	\end{comment}
	
	For $\left|\Delta_{2}\right|$, for each $\sigma\in\left[\tau-\frac{2\pi}{\omega},\tau\right]$
	where $\tau\in\mathbb{R}$, define $\epsilon_{1}\left[\sigma\right],\epsilon_{2}\left[\sigma\right]\in\mathcal{C}\left(\left[-\tau_{d},0\right];\mathcal{Q}\right)$
	where $\mathcal{Q}\subset\mathbb{R}$ is a compact set, recall the
	definition of $J$ and note (\ref{eq:q}), and then it could be seen
	that 
	\begin{align*}
	& \left|J\left(0,\xi_{r}\left[\sigma\right],\epsilon_{1}\left[\sigma\right],N\right)-J\left(0,\xi_{r}\left[\sigma\right],\epsilon_{2}\left[\sigma\right],N\right)\right|\\
	& \leq\left|g_{e}\circ L_{e}\circ\bar{e}_{z1}\left[\sigma\right]-g_{e}\circ L_{e}\circ\bar{e}_{z2}\left[\sigma\right]\right|\\
	& \phantom{\leq}+\left|g_{s}\circ L_{s}\circ\epsilon_{1}\left[\sigma\right]-g_{s}\circ L_{s}\circ\epsilon_{2}\left[\sigma\right]\right|,
	\end{align*}
	where $\bar{e}_{z1}\left[\sigma\right]\triangleq h_{z}\circ M\left(\xi_{r}\left[\sigma\right],\epsilon_{1}\left[\sigma\right],N\right)$
	and $\bar{e}_{z2}\left[\sigma\right]\triangleq h_{z}\circ M\left(\xi_{r}\left[\sigma\right],\epsilon_{2}\left[\sigma\right],N\right)$.
	Notice the facts that $L_{e}\circ\bar{e}_{z1}\left[\sigma\right]-L_{e}\circ\bar{e}_{z2}\left[\sigma\right]\leq\max_{r\in\left[-T_{l},0\right]}\left|\bar{e}_{z1}\left(\sigma+r\right)-\bar{e}_{z2}\left(\sigma+r\right)\right|$
	by the definition of $L_{e}$ in (\ref{eq:qe}) and that $L_{s}\circ\epsilon_{1}\left[\sigma\right]-L_{s}\circ\epsilon_{2}\left[\sigma\right]\leq\max_{r\in\left[-T_{l},0\right]}\left|\epsilon_{1}\left(\sigma+r\right)-\epsilon_{2}\left(\sigma+r\right)\right|$
	for $\tau_{\phi}\leq T_{l}$ %
	\begin{comment}
	Private remark: Is this really justifiable? Also, is this necessary?
	Can we just consider $r\in\left[-\tau_{d},0\right]$ instead? 
	\end{comment}
	by the definition of $L_{s}$ (\ref{eq:qs}). Then, due to the smoothness
	of $g_{e}$, $g_{s}$, $h_{z}$ and $M$, it can be seen that there
	exists $C_{M}>0$ such that 
	\begin{align*}
	& \left|J\left(0,\xi_{r}\left[\sigma\right],\epsilon_{1}\left[\sigma\right],N\right)-J\left(0,\xi_{r}\left[\sigma\right],\epsilon_{2}\left[\sigma\right],N\right)\right|\\
	& \leq C_{M}\max_{r\in\left[-T_{l},0\right]}\left|\epsilon_{1}\left(\sigma+r\right)-\epsilon_{2}\left(\sigma+r\right)\right|.
	\end{align*}
	Now, for each $\sigma\in\left[\tau-\frac{2\pi}{\omega},\tau\right]$
	where $\tau\geq-T_{l}$, define $\bar{\epsilon}_{\sigma}\left(r\right)=\hat{\epsilon}\left(\sigma\right)+as\left(\sigma\right)$
	for all $r\in\mathbb{R}$, then by definition $\bar{J}\left(\hat{\epsilon}\left(\sigma\right)+as\left(\sigma\right),N,\sigma\right)=J\left(0,\xi_{r}\left[\sigma\right],\bar{\epsilon}_{\sigma}\left[\sigma\right],N\right)$.
	Also, since by assumption $\tilde{\epsilon}\left[\sigma\right],\epsilon^{*}\left[N,\sigma\right]\in\mathcal{C}\left(\left[-\tau_{d},0\right];\mathcal{Q}\right)$,
	it can be seen that $\epsilon\left[\sigma\right],\bar{\epsilon}_{\sigma}\left[\sigma\right]\in\mathcal{C}\left(\left[-\tau_{d},0\right];\mathcal{Q}\right)$
	as well. Therefore, 
	\begin{align*}
	& \left|J\left(0,\xi_{r}\left[\sigma\right],\epsilon\left[\sigma\right],N\right)-\bar{J}\left(\hat{\epsilon}\left(\sigma\right)+as\left(\sigma\right),N,\sigma\right)\right|\\
	& =\left|J\left(0,\xi_{r}\left[\sigma\right],\epsilon\left[\sigma\right],N\right)-J\left(0,\xi_{r}\left[\sigma\right],\bar{\epsilon}_{\sigma}\left[\sigma\right],N\right)\right|\\
	& \leq C_{M}\max_{r\in\left[-T_{l},0\right]}\left|\epsilon\left(\sigma+r\right)-\bar{\epsilon}_{\sigma}\left(\sigma+r\right)\right|\\
	& =C_{M}\max_{r\in\left[-T_{l},0\right]}\left|\epsilon\left(\sigma+r\right)-\hat{\epsilon}\left(\sigma\right)-as\left(\sigma\right)\right|\\
	& \leq C_{M}\max_{r\in\left[-T_{l},0\right]}\left|\hat{\epsilon}\left(\sigma+r\right)-\hat{\epsilon}\left(\sigma\right)\right|\\
	& \phantom{\leq}+C_{M}\max_{r\in\left[-T_{l},0\right]}\left|a\sin\left(\omega\left(\sigma+r\right)\right)-a\sin\left(\omega\left(\sigma-\tau_{\phi}\right)\right)\right|
	\end{align*}
	for all $\sigma\in\left[\tau-\frac{2\pi}{\omega},\tau\right]$ where
	$\tau\geq0$. Notice that 
	\begin{align*}
	& \max_{r\in\left[-T_{l},0\right]}\left|\hat{\epsilon}\left(\sigma+r\right)-\hat{\epsilon}\left(\sigma\right)\right|\\
	& \leq T_{l}\max_{r\in\left[\sigma-T_{l},\sigma\right]}\left|\frac{d\hat{\epsilon}}{d\tau}\left(r\right)\right|\\
	& \leq a\omega\delta T_{l}c_{J2}
	\end{align*}
	holds for all $\sigma\in\left[\tau-\frac{2\pi}{\omega},\tau\right]$
	where $\tau\geq\tau_{d}$, %
	\begin{comment}
	Private remark: $\tilde{x}\left(r\right),\tilde{\epsilon}\left(r\right)$
	defined on $r\geq-\tau_{d}=-T_{l}-\frac{2\pi}{\omega}$, so $\frac{\omega}{2\pi}\int_{r-\frac{2\pi}{\omega}}^{r}\left|J\left(\tilde{x}\left[r\right],\epsilon\left[r\right],\xi_{r}\left[r\right]\right)\right|dr$
	defined on $r\geq0$, so $2a\omega\delta T_{l}\max_{r\in\left[\sigma-T_{l},\sigma\right]}\frac{\omega}{2\pi}\int_{r-\frac{2\pi}{\omega}}^{r}\left|J\left(\tilde{x}\left[r\right],\epsilon\left[r\right],\xi_{r}\left[r\right]\right)\right|dr$
	is defined on $\sigma\geq T_{l}$, so we need $\tau\geq T_{l}+\frac{2\pi}{\omega}$. 
	\end{comment}
	and 
	\begin{align*}
	& \max_{r\in\left[-T_{l},0\right]}\left|a\sin\left(\omega\left(\sigma+r\right)\right)-a\sin\left(\omega\left(\sigma-\tau_{\phi}\right)\right)\right|\\
	& =\max_{r\in\left[-T_{l},0\right]}\left|\int_{\sigma-\tau_{\phi}}^{\sigma+r}a\omega\cos\left(\omega\rho\right)d\rho\right|\\
	& \leq a\omega\max_{r\in\left[-T_{l},0\right]}\left|r-\tau_{\phi}\right|
	\end{align*}
	holds for all $\sigma\in\left[\tau-\frac{2\pi}{\omega},\tau\right]$
	where $\tau\in\mathbb{R}$. Therefore, it follows that there exists
	some constant $C_{\Delta_{2}}>0$ such that 
	\[
	\left|\Delta_{2}\right|\leq a\omega C_{\Delta_{2}}
	\]
	for all $\tau\geq\tau_{d}$.
	
	For $\left|\Delta_{3}\right|$, using a similar analysis as before,
	it could be seen that for each $\sigma\in\left[\tau-\frac{2\pi}{\omega},\tau\right]$
	where $\tau\geq0$, 
	\begin{align*}
	& \left|J\left(\tilde{x}\left[\sigma\right],\xi_{r}\left[\sigma\right],\epsilon\left[\sigma\right],N\right)-J\left(0,\xi_{r}\left[\sigma\right],\epsilon\left[\sigma\right],N\right)\right|\\
	& =\left|g_{e}\circ L_{e}\circ h_{z}\left(\tilde{x}\left[\sigma\right]+M\left(\xi_{r}\left[\sigma\right],\epsilon\left[\sigma\right],N\right)\right)\right.\\
	& \phantom{=}\left.-g_{e}\circ L_{e}\circ h_{z}\left(M\left(\xi_{r}\left[\sigma\right],\epsilon\left[\sigma\right],N\right)\right)\right|\\
	& \leq C_{J}\max_{r\in\left[-T_{l},0\right]}\left|\tilde{x}\left(\sigma+r\right)\right|.
	\end{align*}
	\begin{comment}
	calculation procedure: 
	\begin{align*}
	\left|J\left(\tilde{x}\left[\sigma\right],\epsilon\left[\sigma\right],\xi_{r}\left[\sigma\right]\right)-J\left(0,\epsilon\left[\sigma\right],\xi_{r}\left[\sigma\right]\right)\right| & =\left|g_{e}\circ L_{e}\circ h_{az}\left(\tilde{x}\left[\sigma\right]+M\left(\epsilon\left[\sigma\right],\xi_{r}\left[\sigma\right]\right)\right)-g_{e}\circ L_{e}\circ h_{az}\left(M\left(\epsilon\left[\sigma\right],\xi_{r}\left[\sigma\right]\right)\right)\right|\\
	& \leq K_{g_{e}}\left|L_{e}\circ h_{az}\left(\tilde{x}\left[\sigma\right]+M\left(\epsilon\left[\sigma\right],\xi_{r}\left[\sigma\right]\right)\right)-L_{e}\circ h_{az}\left(M\left(\epsilon\left[\sigma\right],\xi_{r}\left[\sigma\right]\right)\right)\right|\\
	& \leq K_{g_{e}}L_{e}\left(h_{az}\left(\tilde{x}\left[\sigma\right]+M\left(\epsilon\left[\sigma\right],\xi_{r}\left[\sigma\right]\right)\right)-h_{az}\left(M\left(\epsilon\left[\sigma\right],\xi_{r}\left[\sigma\right]\right)\right)\right)\\
	& \leq K_{g_{e}}\max_{r\in\left[-T_{l},0\right]}\left|h_{az}\left(\tilde{x}\left[\sigma\right]+M\left(\epsilon\left[\sigma\right],\xi_{r}\left[\sigma\right]\right)\right)-h_{az}\left(M\left(\epsilon\left[\sigma\right],\xi_{r}\left[\sigma\right]\right)\right)\right|\\
	& \leq K_{g_{e}}K_{h_{az}}\max_{r\in\left[-T_{l},0\right]}\left|\tilde{x}\left(\sigma+r\right)\right|\\
	& =C_{J}\max_{r\in\left[-T_{l},0\right]}\left|\tilde{x}\left(\sigma+r\right)\right|
	\end{align*}
	\end{comment}
	Therefore, it follows that there exists some constant $C_{\Delta_{3}}>0$
	such that 
	\[
	\left|\Delta_{3}\right|\leq C_{\Delta_{3}}\left|\tilde{x}\left[\tau\right]\right|
	\]
	for all $\tau\geq0$.
	
	For $\left|\Delta_{4}\right|$, since by Assumption \ref{ass:maximum}
	$\frac{\partial^{2}\bar{J}}{\partial\epsilon^{2}}$ is bounded for
	bounded trajectories of $\tilde{\epsilon}\left(\tau\right),\epsilon^{*}\left(N,\tau\right)$
	for all $\sigma\in\left[\tau-\frac{2\pi}{\omega},\tau\right]$ and
	$\tau\geq-\tau_{d}$, it follows that there exists some constant $C_{\Delta_{4}}>0$
	such that 
	\[
	\left|\Delta_{4}\right|\leq a^{2}C_{\Delta_{4}}
	\]
	for all $\tau\geq-\tau_{d}$.
	
	Lastly, by Assumption \ref{ass:slowly-varying}, 
	\[
	\left|\frac{\partial\epsilon^{*}}{\partial\tau}\right|\leq\frac{k_{\epsilon}}{N}
	\]
	for $\tau\geq0$.
	
	Now, using the Lyapunov-Razumikhin function candidate $V_{\epsilon}\left(\tilde{\epsilon}\right)\triangleq\frac{\tilde{\epsilon}^{2}}{2}$,
	it follows that 
	\begin{align*}
	\frac{dV_{\epsilon}}{d\tau} & =a^{2}\omega\delta\frac{\partial\bar{J}}{\partial\epsilon}\left(\hat{\epsilon}\left(\tau\right),{N,\tau}\right)\tilde{\epsilon}+a\omega\delta\Delta_{1}\tilde{\epsilon}+a\omega\delta\Delta_{2}\tilde{\epsilon}\\
	& \phantom{=}+a\omega\delta\Delta_{3}\tilde{\epsilon}+a\omega\delta\Delta_{4}\tilde{\epsilon}-{\frac{\partial\epsilon^{*}}{\partial\tau}\left(N,\tau\right)}\tilde{\epsilon}.
	\end{align*}
	Substituting the obtained upper bounds into the equation while noting
	$\left|\tilde{\epsilon}\right|\leq\rho_{\epsilon}$ yields 
	\begin{align*}
	\frac{dV_{\epsilon}}{d\tau} & \leq-a^{2}\omega\delta\left(\alpha_{J}\left(\left|\tilde{\epsilon}\right|\right)-\delta C_{\Delta_{1}}\rho_{\epsilon}-\omega C_{\Delta_{2}}\rho_{\epsilon}\vphantom{\frac{\delta C_{\Delta_{1}}k_{\epsilon}}{a^{2}}}\right.\\
	& \phantom{\leq}\left.-\frac{1}{a}C_{\Delta_{3}}\left|\tilde{x}\left[\tau\right]\right|\rho_{\epsilon}-aC_{\Delta_{4}}\rho_{\epsilon}-{\frac{a\delta C_{\Delta_{1}}+k_{\epsilon}}{a^{2}\omega\delta N}}\rho_{\epsilon}\right)\\
	& \leq-a^{2}\omega\delta\left(\alpha_{J}\left(\left|\tilde{\epsilon}\right|\right)-\left(\delta+\omega+a+{\frac{1}{a^{2}\omega\delta N}}\right)C_{\epsilon}\rho_{\epsilon}\right.\\
	& \left.-\frac{1}{a}C_{\Delta_{3}}\rho_{\epsilon}\left|\tilde{x}\left[\tau\right]\right|\vphantom{\frac{1}{a^{2}\omega\delta N}}\right)
	\end{align*}
	for some $C_{\epsilon}>0$. Then, provided that 
	\[
	\left|\tilde{\epsilon}\right|\geq\alpha_{J}^{-1}\left(4\left(\delta+\omega+a+\frac{1}{a^{2}\omega\delta N}\right)C_{\epsilon}\rho_{\epsilon}\right)
	\]
	and 
	\[
	\left|\tilde{\epsilon}\right|\geq\alpha_{J}^{-1}\left(\frac{4}{a}C_{\Delta_{3}}\rho_{\epsilon}\left|\tilde{x}\left[\tau\right]\right|\right),
	\]
	it follows that 
	\[
	\frac{dV_{\epsilon}}{d\tau}\leq-\frac{1}{2}a^{2}\omega\delta\alpha_{J}\left(\left|\tilde{\epsilon}\right|\right)
	\]
	for $\tau\geq\tau_{d}$. Therefore, using an analysis that is similar
	to Theorem 4.19 in \cite{Khalil2002}, it can be shown that Proposition
	2 holds with $\gamma_{\epsilon x}\left(\sigma\right)\triangleq\alpha_{J}^{-1}\left(4C_{\Delta_{3}}\rho_{\epsilon}\sigma\right)$
	and $\gamma_{\epsilon C}\left(\sigma\right)\triangleq\alpha_{J}^{-1}\left(4C_{\epsilon}\rho_{\epsilon}\sigma\right)$.
	
	\subsection{Proof of Proposition 3}
	
	Recall that $\frac{d\tilde{\epsilon}}{d\tau}=\frac{d\hat{\epsilon}}{d\tau}-\frac{\partial\epsilon^{*}}{\partial\tau}$
	and using a result from the Proof of Proposition 1, it can be seen
	that there exists some constant $c_{J3}>0$ such that 
	\[
	\left|\frac{d\hat{\epsilon}}{d\tau}\right|\leq2a\omega\delta c_{J3}
	\]
	for $\tau\geq0$. Moreover, by Assumption \ref{ass:slowly-varying},
	\[
	\left|\frac{\partial\epsilon^{*}}{\partial\tau}\right|\leq\frac{k_{\epsilon}}{N}
	\]
	for $\tau\geq0$. Then it follows that
	
	\begin{align*}
	\left|\frac{d\tilde{\epsilon}}{d\tau}\right| & \leq2a\omega\delta c_{J3}+{\frac{k_{\epsilon}}{N}}\\
	& \leq\left(a\omega\delta+{\frac{1}{N}}\right)c_{\epsilon}
	\end{align*}
	for all $\tau\geq0$ for $c_{\epsilon}\triangleq\max\left\{ 2c_{J3},k_{\epsilon}\right\} $.
	Therefore, 
	\[
	\left|\tilde{\epsilon}\left(\tau\right)\right|\leq\left|\tilde{\epsilon}\left(0\right)\right|+\left(a\omega\delta+{\frac{1}{N}}\right)c_{\epsilon}\tau
	\]
	for all $\tau\geq0$.
	
	\subsection{Proof of Main Theorem}
	
	Assume without loss of generality that $\nu_{\epsilon}<\min\left\{ \rho_{\epsilon}^{0},\underbar{\ensuremath{\epsilon^{*}}},\overline{\epsilon}-\overline{\epsilon^{*}}\right\} $
	where $\underbar{\ensuremath{\epsilon^{*}}}$ and $\overline{\epsilon^{*}}$
	are introduced in Assumption \ref{ass:maximum}. Choose $\rho_{x},\rho_{\epsilon}>0$
	in Propositions 1\textendash 3 to be 
	\begin{align*}
	\rho_{x} & >\beta_{x}\left(\rho_{x}^{0},0\right),\\
	\rho_{\epsilon} & >\beta_{\epsilon}\left(\rho_{\epsilon}^{0}+\rho_{\epsilon}^{+},0\right),
	\end{align*}
	where $\beta_{x},\beta_{\epsilon}\in\mathcal{KL}$ are defined in
	Proposition 1 and 2 respectively, and $\rho_{\epsilon}^{+}\in\left(0,\nu_{\epsilon}^{+}\right)$
	is arbitrary. It is also assumed without loss of generality that $\beta_{x}\left(\sigma,0\right)\geq\sigma$
	and $\beta_{\epsilon}\left(\sigma,0\right)\geq\sigma$ for all $\sigma\geq0$.
	\begin{comment}
	\footnote{This observation follows from the special form of $V_{\epsilon}$
	and the proofs of Theorem 4.19 and Lemma 4.4 in \cite{Khalil2002}. } 
	\end{comment}
	Now, recall $c_{\epsilon}$ in Proposition 3 and $k_{\epsilon}$ in
	Assumption \ref{ass:slowly-varying}, and let $a^{*}=\min\left\{ \gamma_{\epsilon C}^{-1}\left(\min\left\{ \rho_{\epsilon},\nu_{\epsilon}\right\} \right),\nu_{\epsilon}^{+}-\rho_{\epsilon}^{+},\underbar{\ensuremath{\epsilon^{*}}}-\nu_{\epsilon},\overline{\ensuremath{\epsilon}}-\overline{\epsilon^{*}}-\nu_{\epsilon}\right\} $,
	then for any $a\in\left(0,a^{*}\right)$, let $\omega^{*}=\min\left\{ \gamma_{\epsilon C}^{-1}\left(\min\left\{ \rho_{\epsilon},\nu_{\epsilon}\right\} \right)-a,\frac{1}{a}\gamma_{xC}^{-1}\left(\min\left\{ \rho_{x},\nu_{x}\right\} \right),\right.$
	$\left.\frac{1}{a}\gamma_{xC}^{-1}\left(a\gamma_{\epsilon x}\left(\min\left\{ \rho_{\epsilon},\nu_{\epsilon}\right\} \right)\right)\right\} $,
	then for any $\omega\in\left(0,\omega^{*}\right)$, let $\tau_{0}^{*}$
	be such that $\beta_{x}\left(\rho_{x}^{0},\tau_{0}^{*}-\tau_{d}\right)=a\gamma_{\epsilon x}^{-1}\left(\min\left\{ \rho_{\epsilon},\nu_{\epsilon}\right\} \right)$,
	then for any $\tau^{*}>\tau_{0}^{*}$, let $\delta^{*}=\min\left\{ \gamma_{\epsilon C}^{-1}\left(\min\left\{ \rho_{\epsilon},\nu_{\epsilon}\right\} \right)-a-\omega,\frac{\rho_{\epsilon}^{+}}{c_{\epsilon}\tau^{*}a\omega}\right\} $,
	then for any $\delta\in\left(0,\delta^{*}\right)$, let $\tau^{**}$
	be such that $\beta_{\epsilon}\left(\rho_{\epsilon}^{0}+\rho_{\epsilon}^{+},a^{2}\omega\delta\tau^{**}\right)=\nu_{\epsilon}$,
	and let $N^{*}=\max\left\{ \frac{1}{a^{2}\omega\delta\left(\gamma_{\epsilon C}^{-1}\left(\min\left\{ \rho_{\epsilon},\nu_{\epsilon}\right\} \right)-a-\omega-\delta\right)},\frac{1}{\frac{\rho_{\epsilon}^{+}}{c_{\epsilon}\tau^{*}}-a\omega\delta},\right.$
	$\left.\frac{k_{\epsilon}\left(\tau^{*}+\tau^{**}\right)}{\nu_{\epsilon}^{+}-\rho_{\epsilon}^{+}-a}\right\} $,
	then for any $N>N^{*}$, it can be shown that the following inequalities
	hold: 
	\begin{gather}
	a\omega<\gamma_{xC}^{-1}\left(\rho_{x}\right),\label{eq:ineq_1_first}\\
	a\omega<\gamma_{xC}^{-1}\left(a\gamma_{\epsilon x}^{-1}\left(\rho_{\epsilon}\right)\right),\\
	\beta_{x}\left(\rho_{x}^{0},\tau^{*}-\tau_{d}\right)<a\gamma_{\epsilon x}^{-1}\left(\rho_{\epsilon}\right),\\
	a+\omega+\delta+\frac{1}{a^{2}\omega\delta N}<\gamma_{\epsilon C}^{-1}\left(\rho_{\epsilon}\right),\\
	a\omega\delta+\frac{1}{N}<\frac{\rho_{\epsilon}^{+}}{c_{\epsilon}\tau^{*}},\label{eq:ineq_1_last}\\
	a\omega<\gamma_{xC}^{-1}\left(\nu_{x}\right),\label{eq:ineq_2_first}\\
	a\omega<\gamma_{xC}^{-1}\left(a\gamma_{\epsilon x}^{-1}\left(\nu_{\epsilon}\right)\right),\\
	\beta_{x}\left(\rho_{x}^{0},\tau^{*}-\tau_{d}\right)<a\gamma_{\epsilon x}^{-1}\left(\nu_{\epsilon}\right),\\
	a+\omega+\delta+\frac{1}{a^{2}\omega\delta N}<\gamma_{\epsilon C}^{-1}\left(\nu_{\epsilon}\right),\label{eq:ineq_2_last}\\
	\frac{k_{\epsilon}}{N}<\frac{\nu_{\epsilon}^{+}-\rho_{\epsilon}^{+}-a}{\tau^{*}+\tau^{**}}.\label{eq:ineq_3_last}
	\end{gather}
	It can be firstly proved by contradiction that the bounds on the system
	solution required by Propositions 1\textendash 3 indeed hold, i.e.
	$\sup_{\tau\geq0}\left|\tilde{x}\left[\tau\right]\right|\leq\rho_{x}$
	and $\sup_{\tau\geq0}\left|\tilde{\epsilon}\left[\tau\right]\right|\leq\rho_{\epsilon}$.
	Suppose it does not and there exists $\tau_{2}>0$ such that either
	$\left|\tilde{x}\left(\tau_{2}\right)\right|>\rho_{x}$ or $\left|\tilde{\epsilon}\left(\tau_{2}\right)\right|>\rho_{\epsilon}$.
	Due to the continuity of the solution, and notice that $\left|\tilde{x}\left(0\right)\right|\leq\rho_{x}^{0}<\rho_{x}$
	and $\left|\tilde{\epsilon}\left(0\right)\right|\leq\rho_{\epsilon}^{0}<\rho_{\epsilon}$,
	there must exist $\tau_{1}>0$ such that 
	\begin{equation}
	\left|\tilde{x}\left(\tau_{1}\right)\right|=\rho_{x}\text{ or }\left|\tilde{\epsilon}\left(\tau_{1}\right)\right|=\rho_{\epsilon}\label{eq:contra1}
	\end{equation}
	but 
	\[
	\sup_{\tau\in\left[0,\tau_{1}\right]}\left|\tilde{x}\left[\tau\right]\right|\leq\rho_{x}\text{ and }\sup_{\tau\in\left[0,\tau_{1}\right]}\left|\tilde{\epsilon}\left[\tau\right]\right|\leq\rho_{\epsilon}.
	\]
	However, it should be noted that (\ref{eq:prop1})\textendash (\ref{eq:prop3})
	are still valid for $\tau\in\left[0,\tau_{1}\right]$ due to causality.
	Therefore, using (\ref{eq:prop1})\textendash (\ref{eq:ineq_1_last})
	it can be shown that the system solution satisfies 
	\[
	\sup_{\tau\in\left[0,\tau_{1}\right]}\left|\tilde{x}\left[\tau\right]\right|<\rho_{x}\text{ and }\sup_{\tau\in\left[0,\tau_{1}\right]}\left|\tilde{\epsilon}\left[\tau\right]\right|<\rho_{\epsilon},
	\]
	\begin{comment}
	\begin{align*}
	\sup_{\tau\in\left[0,\tau_{1}\right]}\left|\tilde{x}\left[\tau\right]\right| & \leq\max\left\{ \rho_{x}^{0},\beta_{x}\left(\rho_{x}^{0},0\right),\gamma_{xC}\left(a\omega\right)\right\} <\rho_{x}\\
	\sup_{\tau\in\left[0,\tau_{1}\right]}\left|\tilde{\epsilon}\left[\tau\right]\right| & \leq\max\left\{ \rho_{\epsilon}^{0},\rho_{\epsilon}^{0}+\left(a\omega\delta+\frac{1}{N}\right)c_{\epsilon}\tau^{*},\right.\\
	& \beta_{\epsilon}\left(\rho_{\epsilon}^{0}+\left(a\omega\delta+\frac{1}{N}\right)c_{\epsilon}\tau^{*},0\right),\\
	& \gamma_{\epsilon x}\left(\frac{1}{a}\max\left\{ \beta_{x}\left(\rho_{x}^{0},\tau^{*}-\tau_{d}\right),\gamma_{xC}\left(a\omega\right)\right\} \right),\\
	& \left.\gamma_{\epsilon C}\left(a+\omega+\delta+\frac{1}{a^{2}\omega\delta N}\right)\right\} <\rho_{\epsilon}
	\end{align*}
	\end{comment}
	which contradicts (\ref{eq:contra1}). Therefore, the bounds $\sup_{\tau\geq0}\left|\tilde{x}\left[\tau\right]\right|\leq\rho_{x}$
	and $\sup_{\tau\geq0}\left|\tilde{\epsilon}\left[\tau\right]\right|\leq\rho_{\epsilon}$
	for Propositions 1\textendash 3 must hold. Next, using (\ref{eq:prop1})\textendash (\ref{eq:prop3})
	and (\ref{eq:ineq_2_first})\textendash (\ref{eq:ineq_2_last}) it
	can be shown that 
	\[
	\limsup_{\tau\rightarrow\infty}\begin{bmatrix}\left|\tilde{x}\left(\tau\right)\right|\\
	\left|\tilde{\epsilon}\left(\tau\right)\right|
	\end{bmatrix}\leq\begin{bmatrix}\nu_{x}\\
	\nu_{\epsilon}
	\end{bmatrix}.
	\]
	Finally, the rest of the proof is to show that 
	\[
	0<\epsilon\left(\tau\right)<\overline{\ensuremath{\epsilon}}
	\]
	for all $\tau\geq0$. Noting $\nu_{\epsilon}^{+}\leq\hat{\epsilon}\left(0\right)\leq\overline{\epsilon}-\nu_{\epsilon}^{+}$,
	it can be seen that 
	\[
	\nu_{\epsilon}^{+}-\epsilon^{*}\left(N,0\right)\leq\tilde{\epsilon}\left(0\right)\leq\overline{\epsilon}-\nu_{\epsilon}^{+}-\epsilon^{*}\left(N,0\right).
	\]
	Using a result from the Proof of Proposition 3 and (\ref{eq:ineq_1_last}),
	it can be shown that 
	\[
	\left|\tilde{\epsilon}\left(\tau\right)-\tilde{\epsilon}\left(0\right)\right|\leq\rho_{\epsilon}^{+}
	\]
	and therefore 
	\[
	\nu_{\epsilon}^{+}-\rho_{\epsilon}^{+}-\epsilon^{*}\left(N,0\right)\leq\tilde{\epsilon}\left(\tau\right)\leq\overline{\epsilon}-\nu_{\epsilon}^{+}+\rho_{\epsilon}^{+}-\epsilon^{*}\left(N,0\right)
	\]
	for $0\leq\tau\leq\tau^{*}$. Since $\tilde{\epsilon}$ is a scalar
	variable, $\tilde{\epsilon}\left(\tau\right)$ converges monotonically
	towards the ball of radius $\nu_{\epsilon}$ for $\tau\geq\tau^{*}$
	and stays inside the ball for $\tau\geq\tau^{*}+\tau^{**}$. Therefore,
	\[
	\nu_{\epsilon}^{+}-\rho_{\epsilon}^{+}-\epsilon^{*}\left(N,0\right)\leq\tilde{\epsilon}\left(\tau\right)\leq\overline{\epsilon}-\nu_{\epsilon}^{+}+\rho_{\epsilon}^{+}-\epsilon^{*}\left(N,0\right)
	\]
	for $\tau^{*}\leq\tau\leq\tau^{*}+\tau^{**}$ and 
	\[
	\left|\tilde{\epsilon}\left(\tau\right)\right|\leq\nu_{\epsilon}
	\]
	for $\tau\geq\tau^{*}+\tau^{**}$. Then, noting $\epsilon\left(\tau\right)=\tilde{\epsilon}\left(\tau\right)+\epsilon^{*}\left(N,\tau\right)+a\sin\left(\omega\tau\right)$,
	\[
	\left|\frac{\partial\epsilon^{*}}{\partial\tau}\right|\leq\frac{k_{\epsilon}}{N}
	\]
	by Assumption \ref{ass:slowly-varying}, and (\ref{eq:ineq_3_last}),
	it can be seen that 
	\begin{align*}
	\epsilon\left(\tau\right) & \geq\nu_{\epsilon}^{+}-\rho_{\epsilon}^{+}-\frac{k_{\epsilon}}{N}\left(\tau^{*}+\tau^{**}\right)-a\\
	& >\nu_{\epsilon}^{+}-\rho_{\epsilon}^{+}-\left(\nu_{\epsilon}^{+}-\rho_{\epsilon}^{+}-a\right)-a\\
	& =0
	\end{align*}
	and 
	\begin{align*}
	\epsilon\left(\tau\right) & \leq\overline{\epsilon}-\nu_{\epsilon}^{+}+\rho_{\epsilon}^{+}+\frac{k_{\epsilon}}{N}\left(\tau^{*}+\tau^{**}\right)+a\\
	& <\overline{\epsilon}-\nu_{\epsilon}^{+}+\rho_{\epsilon}^{+}+\left(\nu_{\epsilon}^{+}-\rho_{\epsilon}^{+}-a\right)+a\\
	& =\overline{\ensuremath{\epsilon}}
	\end{align*}
	for $0\leq\tau\leq\tau^{*}+\tau^{**}$, and 
	\begin{align*}
	\epsilon\left(\tau\right) & \geq-\nu_{\epsilon}+\underbar{\ensuremath{\epsilon^{*}}}-a\\
	& >0
	\end{align*}
	and 
	\begin{align*}
	\epsilon\left(\tau\right) & \leq\nu_{\epsilon}+\overline{\epsilon^{*}}+a\\
	& <\overline{\ensuremath{\epsilon}}
	\end{align*}
	for $\tau\geq\tau^{*}+\tau^{**}$. It follows that 
	\[
	0<\epsilon\left(\tau\right)<\overline{\ensuremath{\epsilon}}
	\]
	for all $\tau\geq0$, and the theorem is proved.
	
	\bibliographystyle{IEEEtran_patch}
	\bibliography{Reference_v3}

	\begin{IEEEbiography}[{\includegraphics[width=1in,height=1.25in,clip,keepaspectratio]{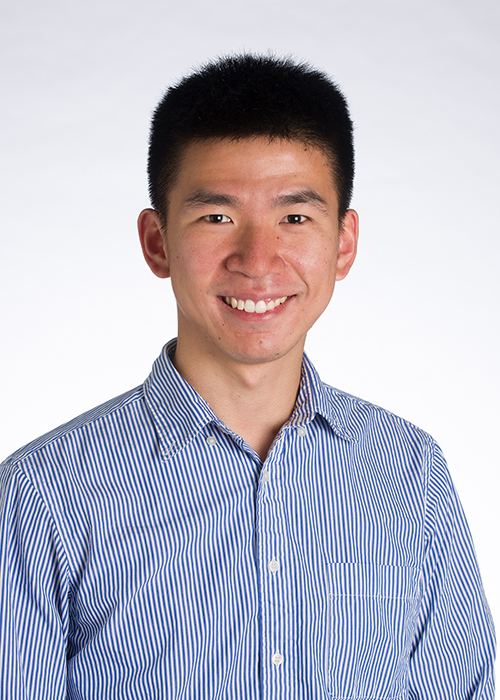}}]{Kaixiang Wang} 
		received his B.E. degree in mechanical engineering from Beijing Institute of Technology and B.E. (Hons) degree in mechatronic systems from the Australian National University in 2014. He is currently in the final stage of his Ph.D. study with the Department of Mechanical Engineering at the University of Melbourne. His thesis investigates the modelling and adaptive scan strategy of atomic force microscopy (AFM). 
	\end{IEEEbiography}
	
	\begin{IEEEbiography}[{\includegraphics[width=1in,height=1.25in,clip,keepaspectratio]{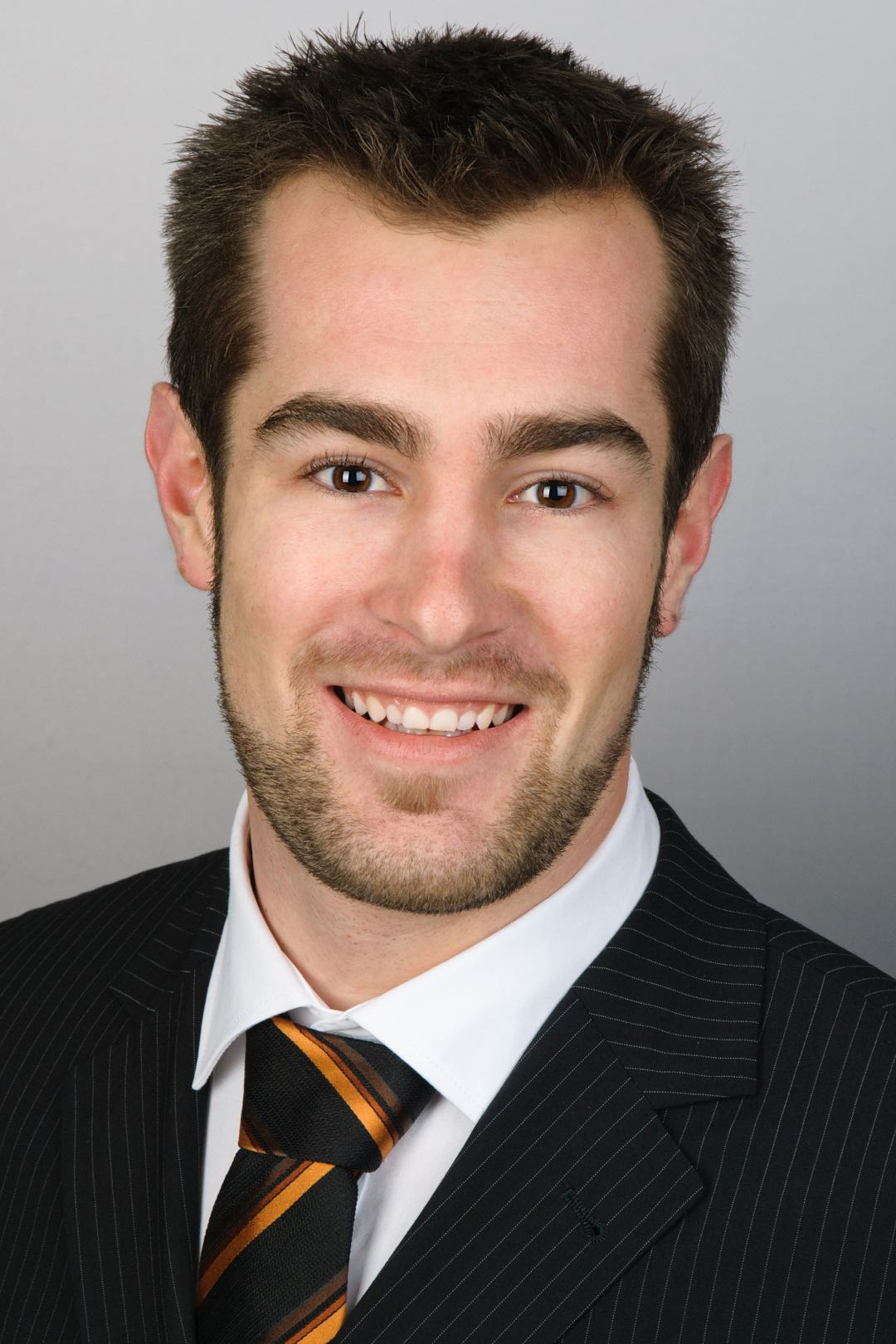}}]{Michael G. Ruppert} 
		(M'14) received the Dipl.-Ing. degree in automation technology in production from the University of Stuttgart, Germany, in 2013 and the Ph.D. degree in electrical engineering from The University of Newcastle, Australia in 2017. As a visiting researcher, he was with The University of Texas at Dallas, USA from 2015 until 2016 and he is now a lecturer at The University of Newcastle. His research interests include the utilization of system theoretic tools for sensing, estimation and control in high-speed and multifrequency atomic force microscopy.
		Dr Ruppert held an Academic Merit Scholarship from the University of Stuttgart, the Baden-W\"{u}rttemberg Scholarship and post-graduate research scholarships with The University of Newcastle and CSIRO, Clayton, VIC, Australia. Dr Ruppert's research has been recognized with the 2013 IEEE/ASME AIM Best Conference Paper Finalist Award, the 2017 University of Newcastle Higher Degree by Research Excellence Award, the 2018 MARSS Best Conference Paper Finalist Award and the 2018 IEEE TCST Outstanding Paper Award.
	\end{IEEEbiography}
	
	\begin{IEEEbiography}[{\includegraphics[width=1in,height=1.25in,clip,keepaspectratio]{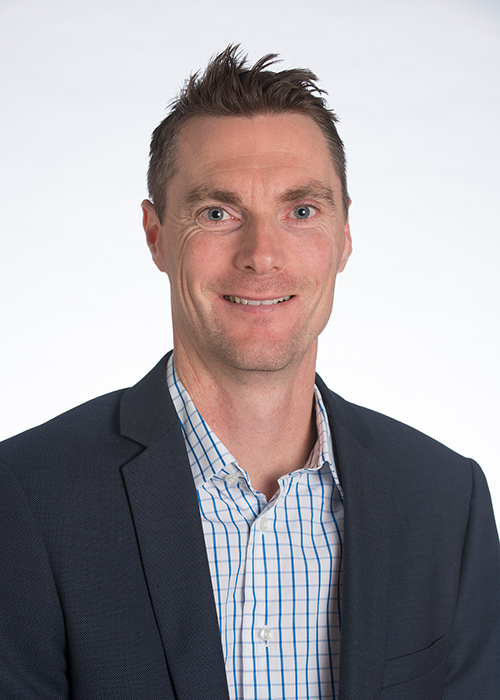}}]{Chris Manzie} 
		(SM'14) is a Professor and Head of Department of Electrical and Electronic Engineering at the University of Melbourne. His research interests are in model-based and model-free control and optimisation, with applications in a range of areas including systems related to energy, transportation and mechatronics and autonomous systems. He received an Future Fellowship from the Australian Research Council for the period 2012-2015. Professor Manzie was also a Visiting Scholar with the University of California, San Diego in 2007 and a Visiteur Scientifique at IFP Energies Nouvelles, Rueil Malmaison in 2012. He is or has been an Associate Editor of IEEE Control Systems Technology, IEEE Mechatronics, Elsevier Control Engineering Practice and ASME Mechatronics.
	\end{IEEEbiography}
	
	\begin{IEEEbiography}[{\includegraphics[width=1in,height=1.25in,clip,keepaspectratio]{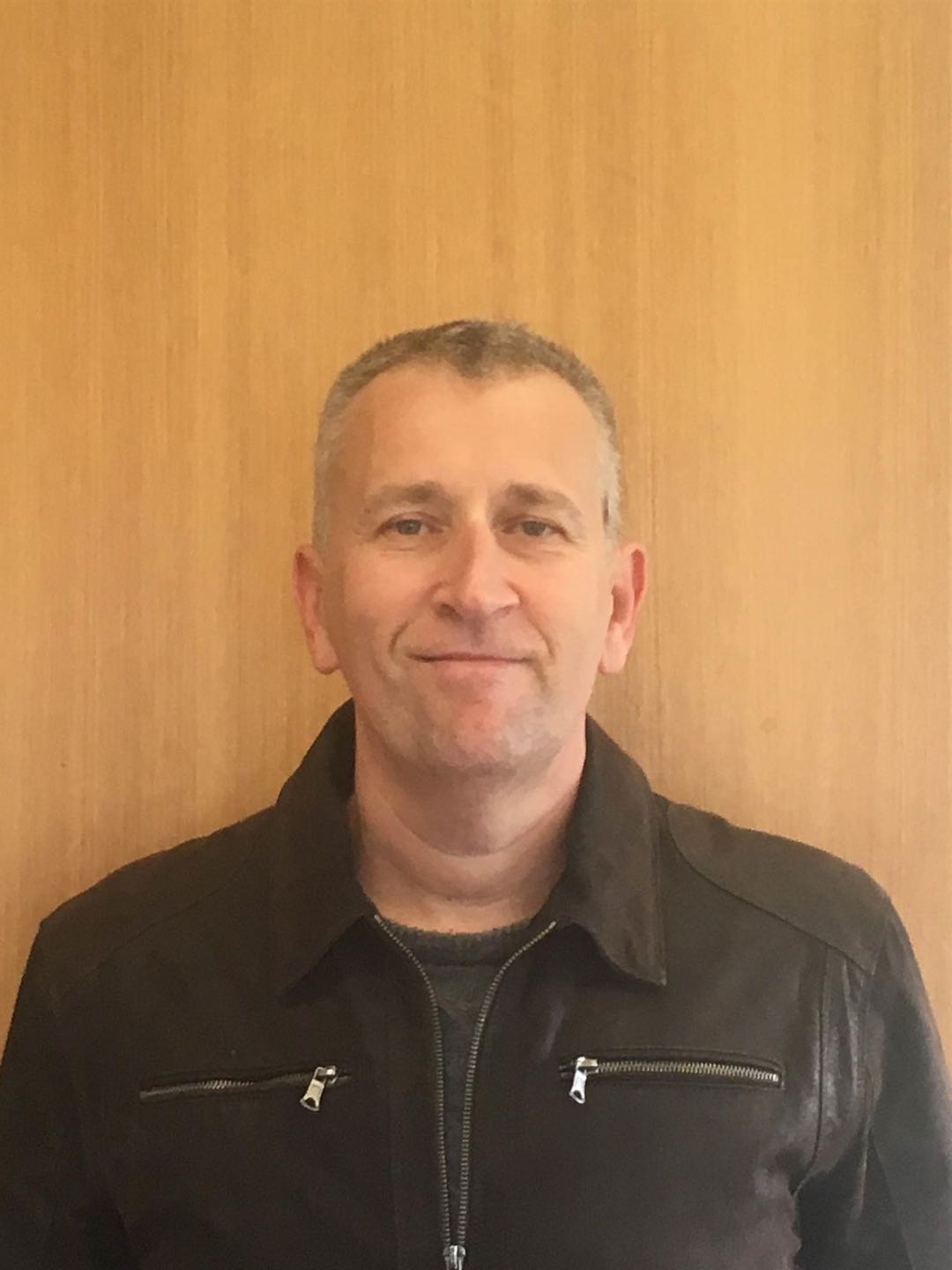}}]{Dragan Nesic} 
		(F'08) is a Professor in the Department of Electrical and Electronic Engineering (DEEE) at The University of Melbourne, Australia. He received his BE degree in Mechanical Engineering from The University of Belgrade, Yugoslavia in 1990, and his Ph.D. degree from Systems Engineering, RSISE, Australian National University, Canberra, Australia in 1997. Since February 1999 he has been with The University of Melbourne. His research interests include networked control systems, reset systems, extremum seeking control, hybrid control systems, and so on. He was a co-receipient of the George S. Axelby Outstanding Paper Award (2018). He was also awarded a Humboldt Research Fellowship (2003) by the Alexander von Humboldt Foundation, an Australian Professorial Fellowship (2004-2009) and Future Fellowship (2010-2014) by the Australian Research Council. He is a Fellow of IEEE. He served as a Distinguished Lecturer of CSS, IEEE and as a Member of the Board of Governors, CSS, IEEE. He also served as an Associate Editor for the journals Automatica, IEEE Transactions on Automatic Control, Systems and Control Letters, European Journal of Control and as a General Co-Chair of IEEE CDC 2017 in Melbourne. Currently, he serves as an Associate Editor for the IEEE Transactions on Control of Network Systems (CONES).
	\end{IEEEbiography}
	
	\begin{IEEEbiography}[{\includegraphics[width=1in,height=1.25in,clip,keepaspectratio]{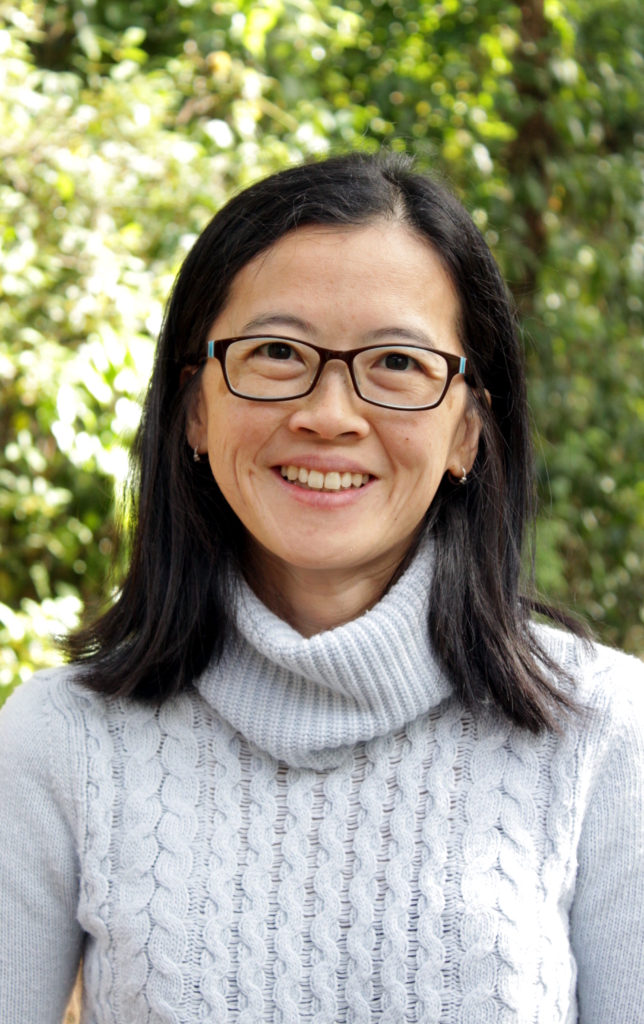}}]{Yuen Kuan Yong} 
		(M'09) is an associate professor at The University of Newcastle, Australia. Her research interests include nanopositioning systems, design and control of micro-cantilevers, atomic force microscopy, and miniature robotics. She was an Australian Research Council DECRA Fellow from 2013 to 2017.  She is also the recipient of the University of Newcastle Vice-Chancellor's Award for Research Excellence in 2014, and the Vice-Chancellor's Award for Research Supervision Excellence in 2017. She is an associate editor for the IEEE/ASME Transactions of Mechatronics.
	\end{IEEEbiography}
	
\end{document}